\documentclass[aps,prd,amsfonts,amsmath,nofootinbib,tightenlines,preprint,showpacs,longbibliography]{revtex4-1}
\pdfoutput=1

\def\be{\begin{equation}}
\def\ee{\end{equation}}
\def\ba{\begin{array}{cccc}}
\def\ea{\end{array}}

\def\tA{\tilde A}
\def\tB{\tilde B}

\usepackage{graphicx}
\graphicspath{ {./Figures/} }

\usepackage{amsmath}
\def\bea#1\eea{\begin{align}#1\end{align}}
\def\blea#1\elea{\begin{subequations}\bea #1\eea\end{subequations}}
\def\bleal#1#2\elea{\begin{subequations}\label{#1}\bea #2\eea\end{subequations}}

\usepackage{array}

\begin{document}

\title{Gravitational back reaction on piecewise linear cosmic string loops}

\author{Jeremy M. Wachter}
\email{Jeremy.Wachter@tufts.edu}
\author{Ken D. Olum}
\email{kdo@cosmos.phy.tufts.edu}
\affiliation{Institute of Cosmology, Department of Physics and Astronomy,\\ 
Tufts University, Medford, MA 02155, USA}

\begin{abstract}
We calculate the metric and affine connection due to a piecewise linear cosmic string loop, and the effect of gravitational back reaction for the Garfinkle-Vachaspati loop with four straight segments. As expected, back reaction reduces the size of the loop, in accord with the energy going into gravitational waves. The ``square'' (maximally symmetric) loop evaporates without changing shape, but for all other loops in this class, the kinks become less sharp and segments between kinks become curved. If the loop is close to the square case, it will evaporate before its kinks are significantly changed; if it is far from square, the opening out of the kinks is much faster than evaporation of the loop.
\end{abstract}


\maketitle

\section{Introduction}\label{sec:intro}

Cosmic strings are a generic prediction of grand unified theories \cite{Jeannerot:2003qv} and may also form in string-theory models of inflation \cite{Sarangi:2002yt,Dvali:2003zj}. They are predicted to exist in ``networks'' of super-horizon strings and oscillating loops. In this work, we analytically solve for the evolution of one class of these loops. While the length scales of loops are astrophysical, string cores are infinitesimally thin, proportional to the inverse of the energy scale of their formation. For this reason, strings are often approximated as one-dimensional objects, and their motion treated via the Nambu action. For a review of cosmic strings, see Ref.~\cite{Vilenkin:2000jqa}.

Cosmic strings are expected to produce observable signatures via gravitational \cite{Damour:2001bk} and electromagnetic \cite{Cai:2012zd} waves, as well as particle production \cite{Berezinsky:2011cp}. Non-observation of gravitational waves \cite{Blanco-Pillado:2013qja,Aasi:2013vna} provides a bound on the dimensionless parameter $G\mu$, where $G$ is Newton's constant and $\mu$ is the mass per unit length of the string. Significant to all forms of emission from cosmic strings are cusps, where the string doubles back on itself and moves with extremely high Lorentz factor, providing a mechanism for observable emissions. Simulations show \cite{Blanco-Pillado:2015ana} that cusps are not present when a loop is formed, but gravitational back reaction may introduce them by smoothing sharp kinks that are present at formation. This process is therefore of great importance in the search for cosmic string signatures.

In this work, we first present a general expression for the metric perturbation, to first order in $G\mu$, due to an arbitrary piecewise linear cosmic string loop (extending the work of Allen and Ottewill \cite{Allen:2000ia} who calculated the metric in the wave zone). We then compute the affine connection due to this metric, which can be used to compute the gravitational back reaction on any piece of the loop. Calculating the back reaction for all but the simplest loops is prohibitively complex, but we analytically calculate such effects on the specific class of rectangular loops (consisting of four straight segments connected by right angle kinks), discussed by Garfinkle and Vachaspati \cite{Garfinkle:1987yw}.

In Sec.~\ref{sec:gen-mp}, we find the general contribution to the metric perturbation and affine connection due to a straight string segment. In Sec.~\ref{sec:spec}, we specialize to the case of rectangular loops. In Sec.~\ref{sec:mod-ws}, we determine how the shape of the string changes as a consequence of back reaction. In Sec.~\ref{sec:eff-brxn}, we provide illustrations and discussions of these changes and other effects of back reaction. Sec.~\ref{sec:conc} concludes and summarizes our work.

We use the metric signature $(-+++)$ and work in units where $\hbar=c=1$.

\section{The metric perturbation of a piecewise linear loop}\label{sec:gen-mp}

We will consider an oscillating string loop with $G\mu\ll 1$, so that the effect of back reaction builds up only after many oscillations. Thus we can work in linearized gravity, where $g_{\alpha\beta}=\eta_{\alpha\beta}+h_{\alpha\beta}$, $g^{\alpha\beta}=\eta^{\alpha\beta}-h^{\alpha\beta}$, and $\left|h_{\alpha\beta}\right|\ll1$. We consider this loop to have always been in oscillation in flat space. We will turn on gravitational effects at some time and analyze the resulting evolution.

We parameterize the string worldsheet by a temporal parameter $\tau$ and a spatial parameter $\sigma$. We will always work in the conformal gauge, with the gauge conditions
\bleal{eqn:gc-conf}
\dot x\cdot x'&=0\,,\\
\dot x^2+x'^2&=0\,,
\elea
where a dot means the derivative with respect to $\tau$ and a prime the derivative with respect to $\sigma$.

When gravitational effects are turned off, the string is moving in flat space. In that case, we may additionally choose $t = x^0=\tau$. Then the equations of motion are just
\be\label{eqn:fs-eom}
\ddot x = x''\,.
\ee
These have the general solution
\be\label{eqn:xAB}
x(t,\sigma) = \frac12\left[A(\tau_-)+B(\tau_+)\right]\,,
\ee
with $\tau_\pm=t\pm\sigma$. $A$ and $B$ are four-vector functions whose tangent vectors $A'$ and $B'$ are null vectors with unit time component.

We now consider the case where the functions $A$ and $B$ are made of linear pieces, so $A'$ and $B'$ each take on a succession of constant values.  The places where each linear piece joins its successor is a kink.  The angles of these kinks could be small, and we could use many pieces to approximate a smooth curve.  Or they could be large and represent actual kinks in the shape of the loop, which could result from previous intersections.

\begin{figure}
\begin{center}
\includegraphics[scale=0.5]{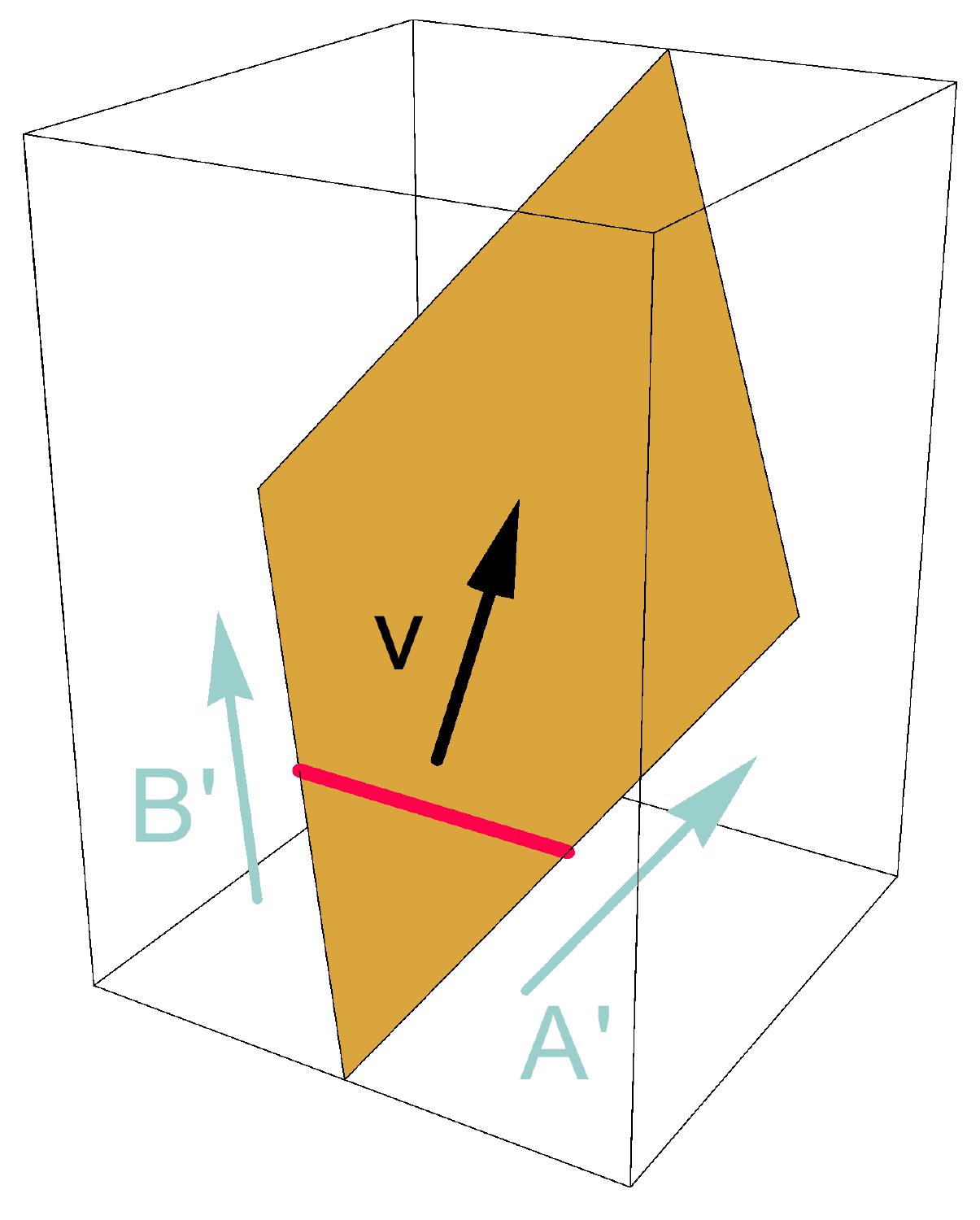}
\caption{An element of the piecewise linear loop, which we call a diamond. The red line on the diamond is a segment of the loop at a fixed time. This segment moves in the direction indicated by $v$, and in so doing sweeps out the diamond. The blue arrows indicate $A'$ and $B'$, which are null, unit length, tangent to the worldsheet, and each parallel to two of the diamond's edges. The diamond shown is a special case where all four edges have the same spatial length. This is true of all diamonds in the Garfinkle-Vachaspati loop we will discuss in Sec.~\ref{sec:spec}.}\label{fig:diamond}
\end{center}
\end{figure}

By Eq.~(\ref{eqn:xAB}), the shape of the string at any fixed time will also be piecewise linear.  The region of the world sheet where a particular linear segment of $A$ combines with a particular segment of $B$ in Eq.~(\ref{eqn:xAB}) is part of a plane in spacetime, traced out by a piece of straight string moving at a constant velocity.  The vectors $A'$ and $B'$, constant over this region, are tangent to that plane.  The line where this segment of $A$ joins its successor is given by fixed $\tau_-$ and thus fixed $A$, while $\tau_+$ varies, so that the line points in the null direction $B'$.  There is a parallel line where this segment of $A$ joins its predecessor.  The two lines where $B$ joins its adjacent segments point in the direction of $A'$. These four lines bound a region of the world sheet as shown in Fig.~\ref{fig:diamond}.  In general the region has the shape of a parallelogram, but we shall refer it as being a \emph{diamond}, which is the case for the rectangular loop we will discuss in Sec.~\ref{sec:spec}.

As we are working in linearized gravity, we may find the metric perturbation in the harmonic gauge due to a given source with stress-energy tensor $T_{\alpha\beta}$ by solving
\be\label{eqn:hab-0}
\Box h_{\alpha\beta}=-16G\pi S_{\alpha\beta}\,,
\ee
where $S_{\alpha\beta} = T_{\alpha\beta} - (1/2)g_{\alpha\beta}T^\gamma_\gamma$ is the trace-reversed stress energy tensor of the source, given by \cite{Quashnock:1990wv,Vilenkin:2000jqa}
\be\label{eqn:def-TRSET}
S^{\alpha\beta}(x)=\mu\int d\tau d\sigma\,\sigma^{\alpha\beta}\delta^{(4)}\left(x-x(\tau,\sigma)\right) = \frac\mu2\int d\tau_- d\tau_+\,\sigma^{\alpha\beta}\delta^{(4)}\left(x-x(\tau_-,\tau_+)\right)
\,,
\ee
with
\be\label{eqn:sab}
\sigma^{\alpha\beta}= \dot x^\alpha\dot x^\beta-x'^\alpha x'^\beta - \frac12\eta^{\alpha\beta}\left[x'^2-\dot x^2\right] = \frac12 \left[A'^\alpha B'^\beta + A'^\beta B'^\alpha - \eta^{\alpha\beta} (A'\cdot B')\right]\,.
\ee

Considering that the diamond edges are traced out by null lines, it may be more convenient to reparameterize the diamond in terms of local null parameters. We choose parameters $u$ and $v$, which are defined separately for each diamond by
\bleal{eqn:uv-def}
u=\tau_+-\tau_+^{(0)}\,,\\
v=\tau_--\tau_-^{(0)}\,,
\elea
where $\tau_\pm^{(0)}$ are the values of $\tau_{\pm}$ at the center of the diamond. By this definition, the range of $u$ is $[-L_B,L_B]$, with $L_B$ the length of the edge of the diamond spanned by $B$. The range of $v$ is defined similarly for $L_A$, and the point $u=v=0$ is the diamond's center.

The diamond worldsheet is now given by
\be\label{eqn:ws-uv}
x(u,v)=\frac12\left[vA'+uB'\right]\,,
\ee
where $A'$ and $B'$ are the unit null tangent vectors to the diamond. The edges of the worldsheet may be traced out by holding either $u = \pm L_B$ or $v = \pm L_A$ fixed and allowing the other null parameter to vary over its full range.

We solve Eq.~(\ref{eqn:hab-0}) using Green's functions,
\be\label{eqn:hab-1}
h_{\alpha\beta}(x)=-16G\pi\sum\int d^4x' S_{\alpha\beta}(x')D_r(x,x')\,,
\ee
for an observation point $x$ and source points $x'$, where the sum is over all contributing source diamonds and $D_r$ is the retarded Green's function,
\be\label{eqn:def-rGf}
D_r(x,x')=-\frac{1}{2\pi}\theta(t-t')\delta\left(\ell^2\right)\,,
\ee
with
\be
\ell =x-x' = \Omega-[vA'+uB']/2\,,
\ee
where $\Omega$ is the vector from the diamond center to the point of observation $x$. Using Eq.~(\ref{eqn:def-TRSET}), the metric perturbation for a single diamond is thus
\be\label{eqn:hab-2}
h_{\alpha\beta}(x)=4G\mu\int d^4x'\,\delta\left(\ell^2(x,x')\right)\int dudv\,\sigma_{\alpha\beta}\delta^{(4)}\left(x'-x(u,v)\right)\,.
\ee
Because the segments move at a constant rate, $\sigma_{\alpha\beta}$ is constant for each diamond and can be taken outside the integral. We integrate over $x'$, consuming the four-dimensional Dirac delta term, and find
\be\label{eqn:hab-3}
h_{\alpha\beta}(x)=4G\mu\sigma_{\alpha\beta}\int dudv\,\delta\left(\ell^2(x,u,v)\right)\,.
\ee

The squared distance from the source to the observation point is
\be\label{eqn:l}
\ell^2=\frac{A'\cdot B'}{2}uv+\Omega^2-(A'\cdot\Omega)v-(B'\cdot\Omega)u\,.
\ee
Setting $\ell^2=0$ gives a hyperbola in $u$ and $v$. This hyperbola has two branches, and we are interested in the one corresponding to the past lightcone. As this corresponds to the intersection of the past lightcone with the string worldsheet, we will refer to it as the \textit{intersection hyperbola}. Solving $\ell^2=0$ for $u$  as a function of $v$ gives
\be\label{eqn:u-f-v}
u(v)=\frac{(A'\cdot\Omega)v-\Omega^2}{(A'\cdot B')v/2-(B'\cdot\Omega)}\,,
\ee
and for $v$ as a function of $u$,
\be\label{eqn:v-f-u}
v(u)=\frac{(B'\cdot\Omega)u-\Omega^2}{(A'\cdot B')u/2-(A'\cdot\Omega)}\,.
\ee
These expressions will be used later in finding the limits of integration. For now, we will rewrite the $\delta$ function via
\be
\delta(\ell^2)=\frac{\delta(v-v(u))}{\left|d\ell^2/dv\right|}\,,
\ee
with
\be
\frac{d\ell^2}{dv}=\frac{A'\cdot B'}{2}u-(A'\cdot\Omega)
\ee
from Eq.~(\ref{eqn:l}). Integrating Eq.~(\ref{eqn:hab-3}) over $v$ then yields
\be\label{eqn:hab-4}
h_{\alpha\beta}=-\frac{8G\mu\sigma_{\alpha\beta}}{A'\cdot B'}\int^{u_+}_{u_-}\frac{du}{2A'\cdot\Omega/(A'\cdot B')-u}=-\frac{8G\mu\sigma_{\alpha\beta}}{A'\cdot B'}\left.\ln\left[\frac{2A'\cdot\Omega}{A'\cdot B'}-u\right]\right|^{u_-}_{u_+}\,.
\ee
Equation~(\ref{eqn:hab-4}) holds also under the exchanges $A'\leftrightarrow B'$, $v\leftrightarrow u$.

The limits of integration, $u_\pm$, are chosen depending on how the intersection hyperbola crosses the diamond. Because the hyperbola must lie on the null past lightcone, it may never be a timelike path, and is only null in the special case where the tip of the lightcone is on the diamond. As such, it may never connect worldsheet edges which are timelike separated, meaning that there are only four kinds of hyperbola which we will need to consider: those connecting two edges where $u=\pm L_B$, those connecting two edges where $v=\pm L_A$, those connecting the two future edges ($u=L_B$ and $v=L_A$) and those connecting two past edges ($u=-L_B$ and $v=-L_A$).

For the case of the intersection hyperbola connecting opposite edges of constant $u$, we choose $u_-=-L_B$ and $u_+=L_B$ to get
\be\label{eqn:hab-X}
h_{\alpha\beta}=-\frac{8G\mu\sigma_{\alpha\beta}}{A'\cdot B'}\ln\left[\frac{-L_BA'\cdot B'/2-A'\cdot\Omega}{L_BA'\cdot B'/2-A'\cdot\Omega}\right]\,.
\ee

The hyperbola which connects opposite edges of constant $v$ has the same solution, but with $A'\leftrightarrow B'$ and $L_B\rightarrow L_A$. For the case of the intersection hyperbola connecting the past edges, we choose $u_-=-L_B$ and find $u_+$ from Eq.~(\ref{eqn:u-f-v}) with $v=-L_A$. We find
\be\label{eqn:hab-A1}
h_{\alpha\beta}=-\frac{8G\mu\sigma_{\alpha\beta}}{A'\cdot B'}\ln\left[\frac{(L_BA'\cdot B'/2+A'\cdot\Omega)(L_AA'\cdot B'/2+B'\cdot\Omega)}{(A'\cdot\Omega)(B'\cdot\Omega)-\Omega^2(A'\cdot B')/2}\right]\,.
\ee

For the case of the intersection hyperbola connecting the future edges, we choose $u_+=L_B$ and find $u_-$ from Eq.~(\ref{eqn:u-f-v}) with $v=L_A$ to get
\be\label{eqn:hab-A2}
h_{\alpha\beta}=-\frac{8G\mu\sigma_{\alpha\beta}}{A'\cdot B'}\ln\left[\frac{(A'\cdot\Omega)(B'\cdot\Omega)-\Omega^2(A'\cdot B')/2}{(L_BA'\cdot B'/2-A'\cdot\Omega)\left(L_AA'\cdot B'\right/2-B'\cdot\Omega)}\right]\,.
\ee
Thus, by knowing the location of the source diamonds and the manner in which the intersection hyperbola crosses each diamond, we may find the metric perturbation at an arbitrary point.

We are interested in the first-order effect of gravitational back reaction on the motion of some string segment, If the unperturbed trajectory of the segment is given by Eq.~(\ref{eqn:xAB}), the perturbation can be found from the equations of motion \cite{Quashnock:1990wv},
\be\label{eqn:eom-0}
x^\lambda_{,uv}=-\Gamma^\lambda_{\alpha\beta}x^\alpha_{,u}x^\beta_{,v} =-\frac14\Gamma^\lambda_{\alpha\beta}B'^\alpha A'^\beta\,,
\ee
where the Christoffel symbol, to first order in the metric perturbation, is given by
\be\label{eqn:CS-def}
\Gamma^\lambda_{\alpha\beta}=\frac12\eta^{\lambda\rho}\left[h_{\beta\rho,\alpha}+h_{\rho\alpha,\beta}-h_{\alpha\beta,\rho}\right]\,.
\ee
Once we have allowed the string to perturb spacetime, we can no longer set $t=\tau$ (although the definitions made in Eq.~(\ref{eqn:uv-def}) will hold always). We note that this perturbation is small and periodic, whereas changes to the string shape due to back reaction will accumulate over many oscillations. Our procedure for finding and distinguishing these secular changes from gauge effects will be to look only at how the worldsheet is modified after $N$ oscillations, where $N \gg 1 \gg N G\mu$, so that secular changes are much larger than oscillatory ones, but still small enough that we don't need to consider second-order effects.

Note that the metric perturbation gets its tensor structure from the $\sigma_{\alpha\beta}$ of the source diamond. But from Eq.~(\ref{eqn:sab}) we can see that any contraction of $\sigma_{\alpha\beta}$ with $A'$ or $B'$ will vanish. Thus if we consider the effect of the source diamond on itself (or any later version of itself), all terms in Eq.~(\ref{eqn:CS-def}) vanish in Eq.~(\ref{eqn:eom-0}) and there is no effect.

The derivatives of the metric perturbation are given by
\be\label{eqn:dhab-X}
h_{\alpha\beta,\gamma} =\frac{8G\mu L_B(A\cdot\Omega)\sigma_{\alpha\beta}A'_\gamma} {(A'\cdot\Omega)^2-L_B^2(A'\cdot B')^2/4}
\ee
for Eq.~(\ref{eqn:hab-X}) connecting constant-$u$ edges,
\bea\label{eqn:dhab-A1}
h_{\alpha\beta,\gamma}=-\frac{8G\mu\sigma_{\alpha\beta}}{A'\cdot B'}\bigg[& \frac{A'_\gamma}{L_BA'\cdot B'/2+A'\cdot\Omega} +\frac{B'_\gamma}{L_AA'\cdot B'/2+B'\cdot\Omega}\\
&- \frac{(A'\cdot\Omega)B'_\gamma+(B'\cdot\Omega)A'_\gamma -(A'\cdot B')\Omega_\gamma} {(A'\cdot\Omega)(B'\cdot\Omega)-\Omega^2(A'\cdot B')/2} \bigg]\nonumber\,.
\eea
for Eq.~(\ref{eqn:hab-A1}) connecting past edges, and
\bea\label{eqn:dhab-A2}
h_{\alpha\beta,\gamma}=-\frac{8G\mu\sigma_{\alpha\beta}}{A'\cdot B'}\bigg[& \frac{A'_\gamma}{L_BA'\cdot B'/2-A'\cdot\Omega} +\frac{B'_\gamma}{L_AA'\cdot B'/2-B'\cdot\Omega}\\
&+ \frac{(A'\cdot\Omega)B'_\gamma+(B'\cdot\Omega)A'_\gamma -(A'\cdot B')\Omega_\gamma} {(A'\cdot\Omega)(B'\cdot\Omega)-\Omega^2(A'\cdot B')/2} \bigg]\nonumber\,.
\eea
for Eq.~(\ref{eqn:hab-A2}) connecting future edges.

Once we have found the acceleration for all diamonds, we find the changes to the null vectors. To do so, we pick a point and follow a null path in $\tau_+$ or $\tau_-$ around the loop worldsheet until we return to the initial point one oscillation in the future. Integrating along this $\tau_+$ path gives the change to $A'$ at some fixed $\tau_-$, while integrating along this $\tau_-$ path gives the change to $B'$ at some fixed $\tau_+$.

\section{Rectangular loops}\label{sec:spec}

\subsection{Geometry}

We now wish to specialize to the rectangular loop of Garfinkle and Vachaspati \cite{Garfinkle:1987yw}, which is simple enough that the entire problem can be solved analytically. In this loop, $A'(\tau_-)$ has some constant value $A'(0)$ for $\tau_-=0\ldots L/2$ and the constant value $-A'(0)$ for $\tau_-=L/2\ldots L$, and $B'(\tau_+)$ behaves similarly, with some angle $2\phi$ between $A'(0)$ and $B'(0)$. By the symmetries of the loop, $\tau_\pm$ should be taken to be modulo $L$.

\begin{figure}
\begin{center}
\includegraphics[scale=1.0]{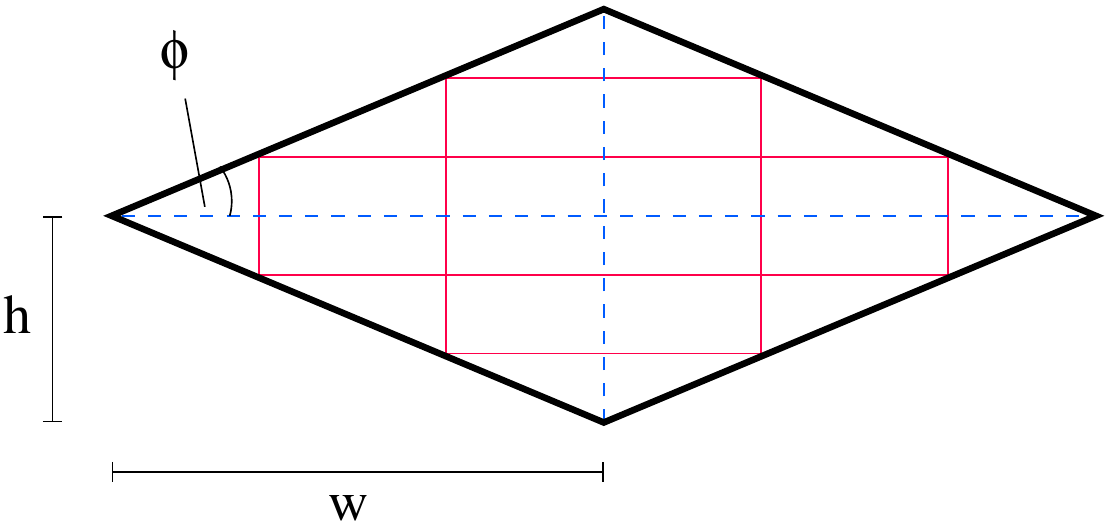}
\caption{The rectangular loop oscillates through all spatial rectangles which may be inscribed within a rhombus whose diagonals (dashed blue) measure $2w$ and $2h$ on the long and short directions respectively. The acute angle of this rhombus is $2\phi$, and so $\tan\phi=h/w$. The solid red rectangles are examples of configurations the loop takes on during its oscillation. The dashed blue lines are additionally degenerate rectangles, also called the double lines, which form when two kinks lie on top of one another.}\label{fig:inscrip}
\end{center}
\end{figure}

We will choose coordinates so that the spatial part of $A'(0)+B'(0)$ points in the $x$ direction and $B'(0)-A'(0)$ points in the $y$ direction. The loop then oscillates through all configurations of a rectangle which may be inscribed within a rhombus of angle $2\phi$, as shown in Fig.~\ref{fig:inscrip}.
Without loss of generality we can choose $0<\phi\leq\pi/4$. The $x$ axis will always be the longer of the rhombus's two diagonals, while the $y$ axis will always be the shorter.

We let $w=\cos\phi$ and $h=\sin\phi$. We then have
\bleal{eqn:AB-spec}
A'^\gamma &= (1, \pm w, \mp h, 0)\,,\\
B'^\gamma &= (1, \pm w, \pm h, 0)\,.
\elea

One oscillation of the loop is the time it takes for the loop to go through all configurations shown in Fig.~\ref{fig:inscrip}, for example from the double line on the $x$ axis to the double line on the $y$ axis and back again. As this loop lies entirely in a plane, it is guaranteed to self-intersect, and therefore annihilate, in this case when it reaches the double-line configurations. But we will take the loop to be infinitesimally thin and not to interact with itself, except in one case below where to avoid a divergence we will need to say that the two parts of the loop pass by each other by some infinitesimal distance.

We will choose $L=4$. Some particulars of the loop's geometry are that
\begin{itemize}
\item $L_A=L_B=1$, so each diamond is a rhombus.
\item The double line on the $x$ axis extends $\pm w$.
\item The double line on the $y$ axis extends $\pm h$.
\item The speed of a segment moving the $x$ direction is $w$, and moving in the $y$ direction is $h$.
\item The period of oscillation is $T=2$.
\end{itemize}

Consider a diamond swept out by a string segment moving in the positive $x$ direction in a rectangular loop. We refer to such a diamond as a {\it $+x$ diamond}. Let the loop be centered on the origin and let the segment pass through the origin at some time $t_0$, so that the diamond center is is $(t_0,0,0)$, its futuremost point is at $(t_0+1,w,0)$ and its pastmost point at $(t_0-1,-w,0)$. We will solve for the effect of back reaction on the $+x$ diamond, and we can then recover the effect on other diamonds as follows.
\begin{itemize}
\item For the $-x$ diamond, exchange $x\leftrightarrow-x$.
\item For the $+y$ diamond, exchange $x\leftrightarrow y$ and $h\leftrightarrow w$.
\item For the $-y$ diamond, exchange $x\leftrightarrow-y$ and $h\leftrightarrow w$.
\end{itemize}
Any of the above may be combined with a redefinition of $t_0$ to change the diamond's position in time.

For the $+x$ diamond, the null vectors are
\bleal{eqn:x+ab}
A'^\gamma&=(1,w,-h,0)\,,\\
B'^\gamma&=(1,w, h,0)\,,
\elea
and we have
\be
\sigma_{\alpha\beta} = \left(\ba w^2&-w&0&0\\-w&1&0&0\\0&0&0&0\\0&0&0&h^2\ea\right)\,.
\ee

\begin{figure}
\begin{center}
\includegraphics[scale=0.6]{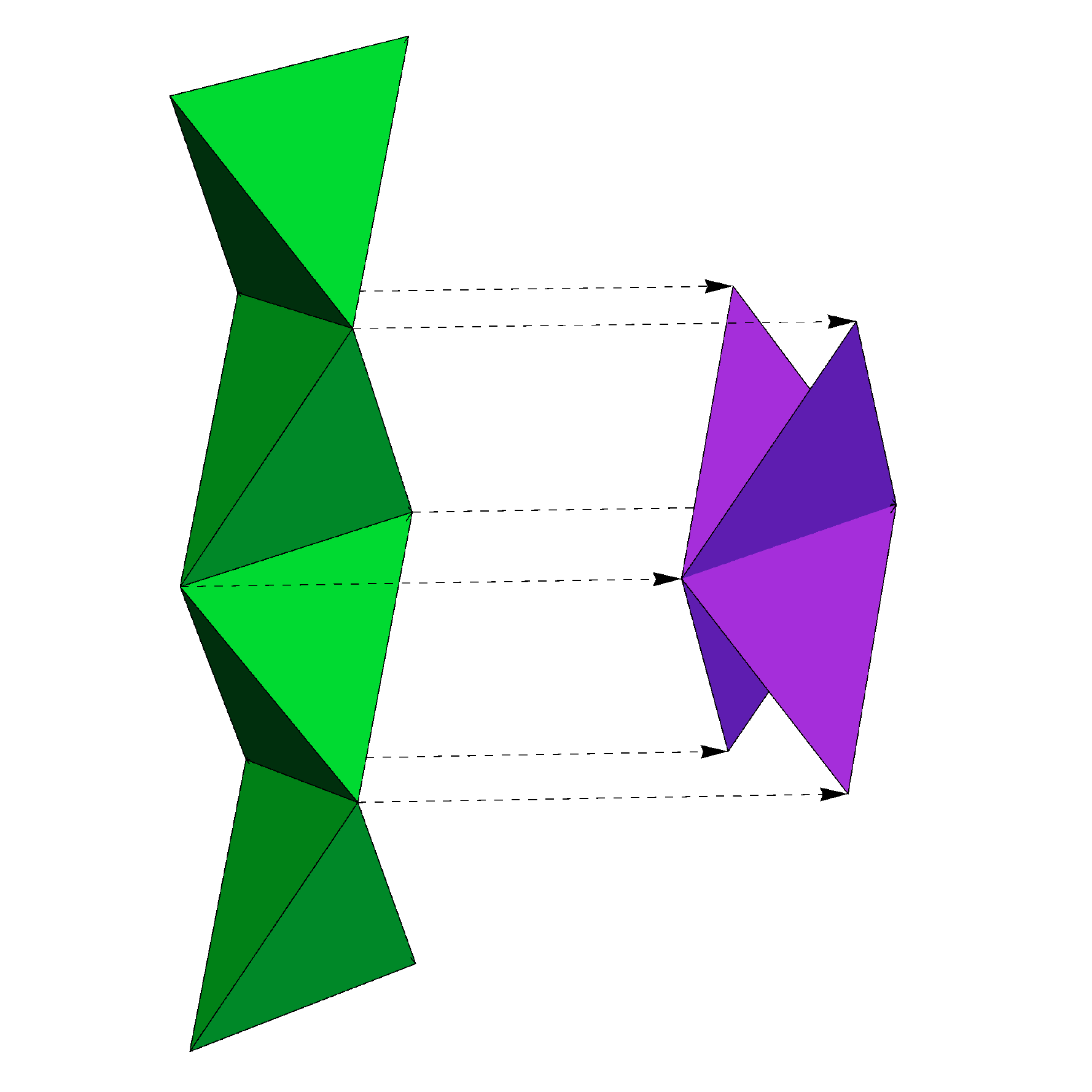}
\caption{On the left is the worldsheet of a loop with $\phi=\pi/6$ over two oscillations, starting from the double-line configuration. Up is forward in time, with the transverse plane being the plane in which the loop is oscillating. By slicing the worldsheet in this transverse plane, we recover all possible configurations of the loop, such as the ones shown (in red and blue) in Fig.~\ref{fig:inscrip}. On the right is an example of a butterfly whose $t_0$ is at the midpoint of the temporal range of the worldsheet.}\label{fig:ws-bfly}
\end{center}
\end{figure}

Another object which we will frequently discuss in the context of the rectangular loop is the combination of a diamond and its spatial mirror image, which we call a {\it butterfly}. A $+x$ and $-x$ diamond with the same $t_0$ taken together are an {\it x butterfly}, while $\pm y$ diamonds with common $t_0$ together make a {\it y butterfly}. The futuremost edges of an $x$ butterfly meet the pastmost edges of the $y$ butterfly whose $t_0$ is greater by $1$. Similarly, the pastmost edges of an $x$ butterfly meet the futuremost edges of the $y$ butterfly whose $t_0$ is less by $1$. Figure~\ref{fig:ws-bfly} illustrates the general appearance of a rectangular loop's worldsheet, and shows how we extract a butterfly from that worldsheet.

\subsection{Metric perturbations}\label{ssec:src-px}

How does the backward lightcone of some observation point intersect the worldsheet of a rectangular loop?  Because the loop and the lightcone are both continuous objects, the total intersection of the lightcone with the worldsheet must be a closed path.  Which diamonds does it cross and how does it cross them?

\begin{figure}
\begin{center}
\includegraphics[scale=0.35]{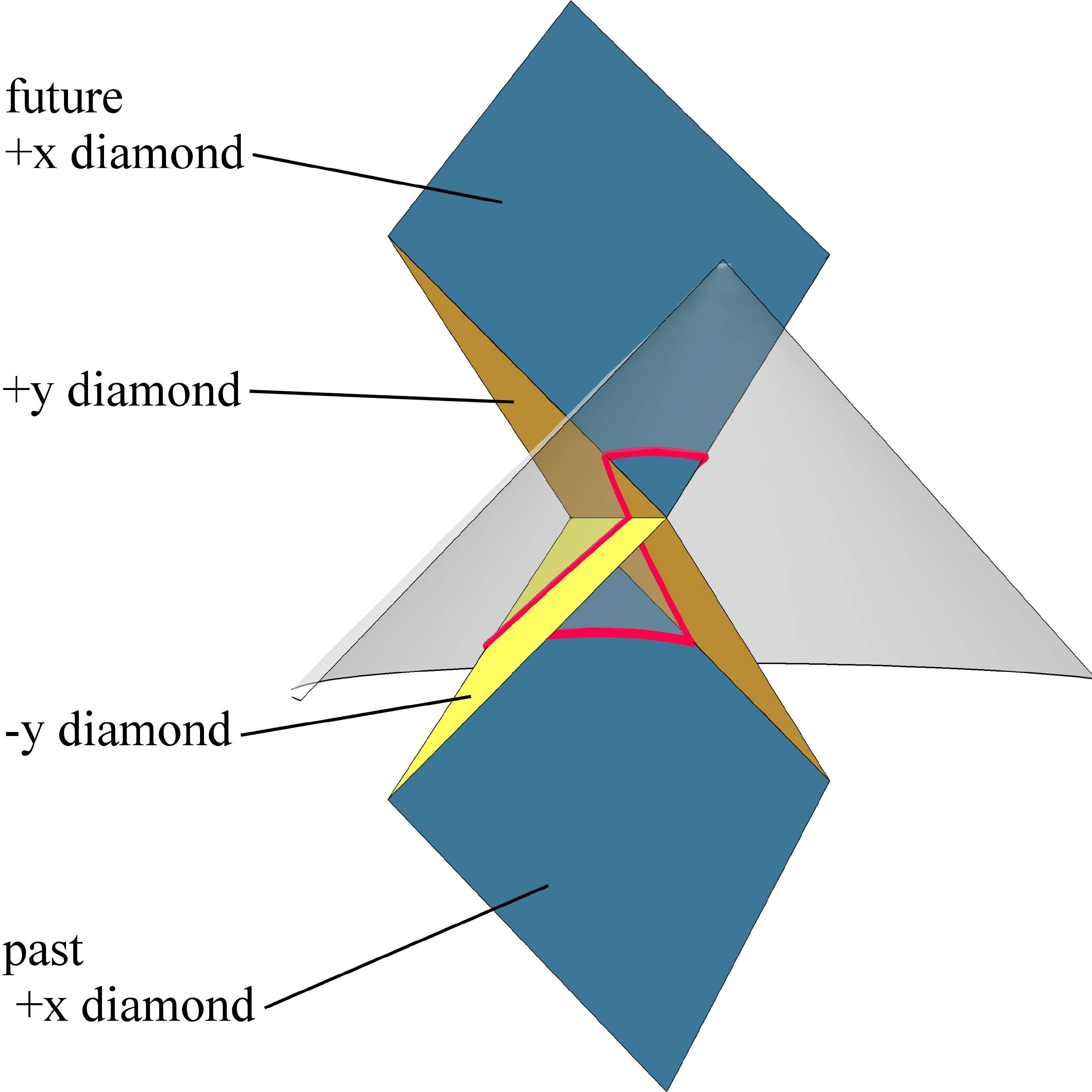}
\caption{The thick red line shows the intersection of the lightcone with the loop worldsheet for a figure-eight intersection. The intersection line passes through two parallel diamonds plus the butterfly in between them for a total of four unique diamonds. We have indicated the diamonds crossed as described in the text.}\label{fig:pdi}
\end{center}
\end{figure}

First suppose that the intersection path connects a past and a future
edge of some diamond.  Suppose, for example, that the path crosses a
$+y$ diamond from the edge it shares with the past $+x$ diamond to the
edge it shares with the future $+x$ diamond, as shown in
Fig.~\ref{fig:pdi}. This $+y$ diamond intersects a $-y$ diamond along
their common centerline.  To get from a past edge to a future edge,
the intersection path must cross this centerline.  (It may only cross
once, it has to get from one side to the other, and no hyperbola can
cross a line three times or more.)  At the place it crosses, it must
also cross part of the intersection path lying on the $-y$ diamond,
which thus must also run from the past $+x$ diamond to the future $+x$
diamond.  (It cannot enter the past $-x$ diamond, because all
interior points on that diamond are in the chronological past of all
points on the future $+x$ diamond, and no two points on the
intersection path can be timelike separated.  The same applies to the
future $-x$ diamond.)

Now the two crossings into the future $+x$ diamond must be
connected to each other via a hyperbola, and so must the two crossings
into the past $+x$ diamond.  Thus we have a closed loop made up
of four diamonds, two $y$ diamonds of the same butterfly and the $+x$
diamonds in its past and future.  We call this a \textit{figure-eight}
intersection.

\begin{figure}
\begin{center}
\includegraphics[scale=0.35]{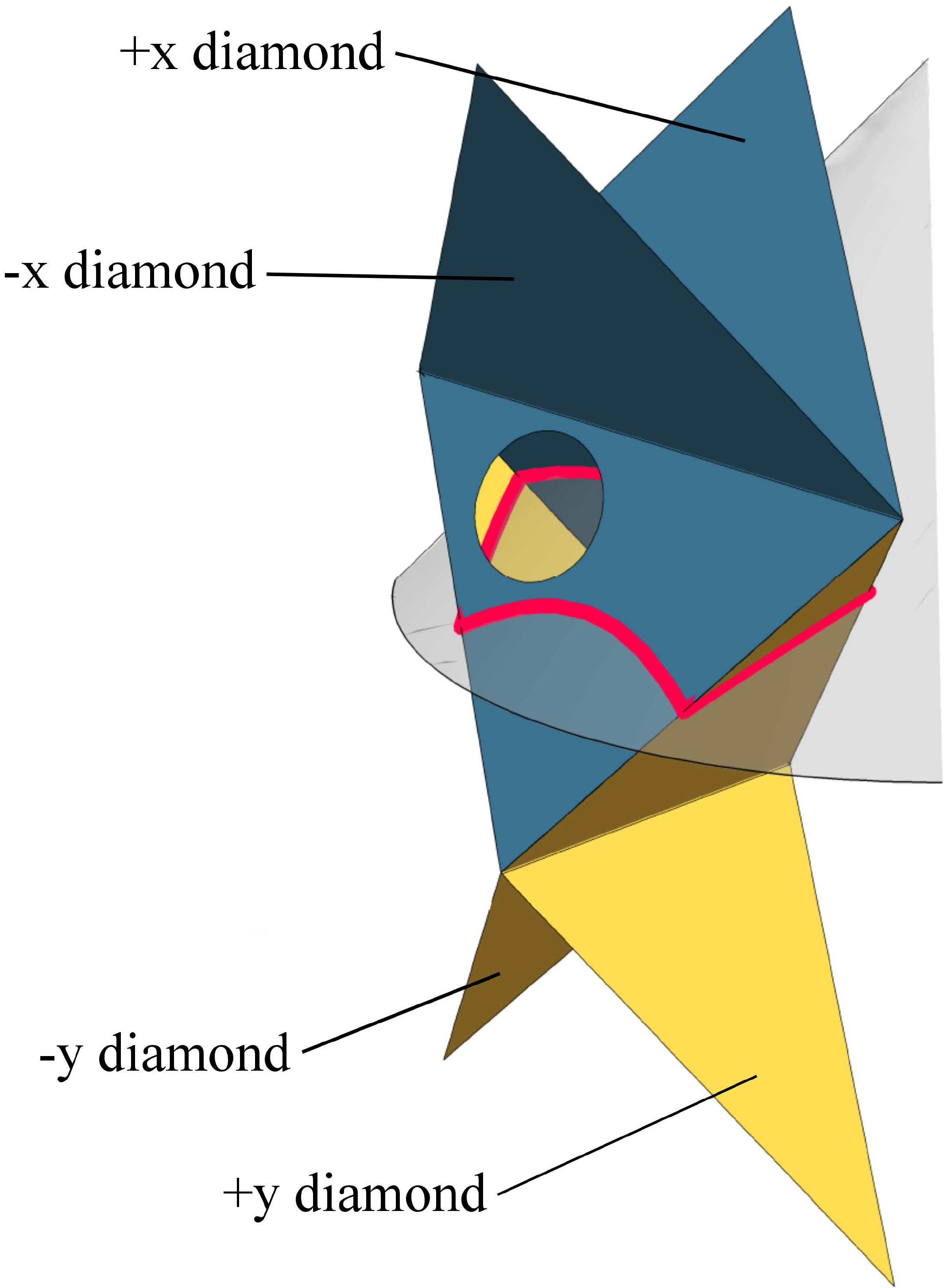}
\caption{The thick red line shows the intersection of the lightcone with the loop worldsheet for a ring intersection. The intersection line passes through two subsequent butterflies for a total of four unique diamonds. We have indicated the diamonds crossed as described in the text.}\label{fig:tbi}
\end{center}
\end{figure}

The only other possibility is that the intersection path never connects a past and a future edge of the same diamond, but only crosses from past to past and from future to future.  Suppose the path connects two past edges of a $+x$ diamond.  At these two places it comes in across the future edges of the $-y$ and $+y$ diamonds.  It must leave these diamonds across their other future edges, which connect to the same $-x$ diamond.  We thus have a closed loop crossing both diamonds of an $x$ butterfly and both diamonds of the prior $y$ butterfly, as shown in Fig.~\ref{fig:tbi}.  We call this a \textit{ring} intersection, although it's possible that the parts of the ring lying on the $x$ butterfly may cross each other at the double line and then cross back again.

We will now calculate the metric perturbation at an observation point $(t,x,y,z)$ whose backward light cone yields either the specific kind of figure-eight intersection or the specific kind of ring intersection described above.  Once we find those effects, we may recover the perturbation for any intersection through the appropriate transpositions and substitutions. 

\begin{figure}
\begin{center}
\includegraphics[scale=0.5]{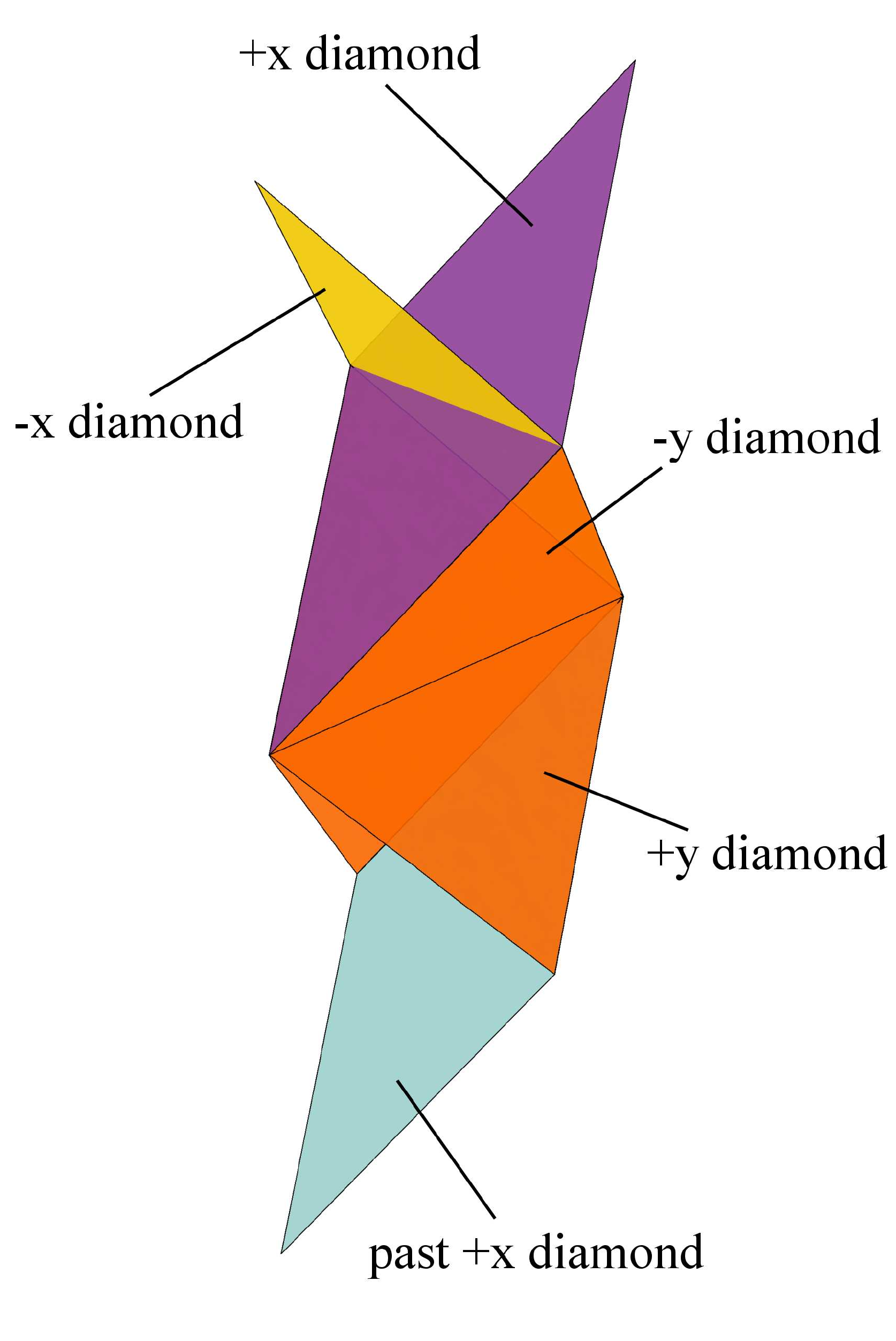}
\caption{The worldsheet of a rectangular loop obtained by combining Figs.~\ref{fig:pdi}~and~\ref{fig:tbi}.}\label{fig:ex-ws}
\end{center}
\end{figure}

We will eventually take the observation point to the worldsheet of a
$+x$ diamond to study how back reaction affects that diamond. When
constrained to that worldsheet, the observation point will only ever
see figure-eight and ring intersections of the specific sort we have
described above, where the same $y$ butterfly is involved in
both. There are only five diamonds total that we need to consider when
finding the effect of back reaction on such a point: an $x$ butterfly
containing the observation point, the $y$ butterfly below the $x$
butterfly, and the $+x$ diamond below the $y$ butterfly. We will call
the components of the $x$ butterfly simply the $+x$ diamond (though it
is the same one we called the ``future $+x$ diamond'' earlier) and the
$-x$ diamond, while the the $+x$ diamond below the $y$ butterfly we
will call the past $+x$ diamond.  Putting
Figs.~\ref{fig:pdi}~and~\ref{fig:tbi} together and color-coding them
for later convenience produces Fig.~\ref{fig:ex-ws}.

We first consider the perturbations associated with the figure-eight intersection. We assume the center of the $x$ butterfly is at $t=0$. Using Eqs.~(\ref{eqn:hab-A1},\ref{eqn:x+ab}), we find
\be\label{eqn:hab-s}
h_{\alpha\beta}^{(+x)}=\frac{4G\mu}{h^2}\sigma_{\alpha\beta}^{(+x)}\ln\left[\frac{(h^2+t-xw)^2-y^2h^2}{(x-tw)^2+h^2z^2}\right]\,.
\ee
The $y$ diamonds, whose centers are at $t=-1$, have perturbations given by Eq.~(\ref{eqn:hab-X}) and its $A'\leftrightarrow B'$ version.
\bleal{eqn:hab-p-2}
h_{\alpha\beta}^{(+y)}&=\frac{4G\mu}{w^2}\sigma_{\alpha\beta}^{(+y)}\ln\left[\frac{1+t-xw-yh+w^2}{1+t-xw-yh-w^2}\right]\,,\\
h_{\alpha\beta}^{(-y)}&=\frac{4G\mu}{w^2}\sigma_{\alpha\beta}^{(-y)}\ln\left[\frac{1+t-xw+yh+w^2}{1+t-xw+yh-w^2}\right]\,.
\elea
For the past $+x$ diamond, which has its center at $t=-2$, equations~(\ref{eqn:hab-A2},\ref{eqn:x+ab}) give
\be\label{eqn:hab-ps}
h_{\alpha\beta}^{(past~+x)}=\frac{4G\mu}{h^2}\sigma_{\alpha\beta}^{(+x)}\ln\left[\frac{(x-tw-2w)^2+h^2z^2}{(h^2-t-2+xw)^2-y^2h^2}\right]\,,
\ee

For the ring intersection, we say that the $y$ butterfly is the same as the one involved in the figure-eight intersection. However, the nature the crossings have changed; we must now use Eq.~(\ref{eqn:hab-A2}), and find
\bleal{eqn:hab-p-1}
h_{\alpha\beta}^{(+y)}&=\frac{4G\mu}{w^2}\sigma_{\alpha\beta}^{(+y)}\ln\left[\frac{(y-th-h)^2+w^2z^2}{(w^2-t-1+yh)^2-x^2w^2}\right]\,,\\
h_{\alpha\beta}^{(-y)}&=\frac{4G\mu}{w^2}\sigma_{\alpha\beta}^{(-y)}\ln\left[\frac{(y+th+h)^2+w^2z^2}{(w^2-t-1-yh)^2-x^2w^2}\right]\,.
\elea
The futuremost $+x$ diamond from the figure-eight intersection is also involved in this ring intersection, and has the same form. The pastmost $+x$ diamond is not involved in the ring intersection, and we must instead consider the perturbation due to a $-x$ diamond whose center is at $t=0$. Using Eq.~(\ref{eqn:hab-A1}),
\be
h_{\alpha\beta}^{(-x)}=\frac{4G\mu}{h^2}\sigma_{\alpha\beta}^{(-x)}\ln\left[\frac{(h^2+t+xw)^2-y^2h^2}{(x+tw)^2+h^2z^2}\right]\,.
\ee

\section{The modified worldsheet}\label{sec:mod-ws}

\subsection{Acceleration}

We are now equipped to find the four-acceleration felt at an arbitrary point due to the loop worldsheet. All accelerations we find from Eq.~(\ref{eqn:eom-0}). As we will eventually take the observation point to lie on the worldsheet of a $+x$ diamond, we take it now to be on a string moving transversely at speed $w$ in the $+x$ direction. As such, its null vectors are identical to that of the $+x$ diamond. 

The acceleration of the string is found by contracting the metric derivatives of Eq.~(\ref{eqn:dhab-X}-\ref{eqn:dhab-A2}) with $A'$ and $B'$, according to Eq.~(\ref{eqn:eom-0}). This means that at least one of $A'$ and $B'$ must be contracted with $\sigma_{\alpha\beta}$. For the $+x$ and past $+x$ diamonds, this contraction vanishes, as discussed earlier.

The $-y$ diamond has the same $A'$ as the observation string. Thus its $\sigma_{\alpha\beta}$ can only be contracted with $B'$, which leaves $A'$ to contract with the direction of differentiation. For a figure-eight intersection, the metric derivative is given by Eq.~(\ref{eqn:dhab-X}), which vanishes on contraction with $A'^\gamma$. The $+y$ diamond is symmetrical and its contribution likewise vanishes, leaving no back reaction effect at all for a figure-eight intersection.

When we consider a ring intersection, the effect of the $-y$ diamond is instead given by Eq.~(\ref{eqn:dhab-A2}). Contracting with $A'^\gamma$ leaves only the middle term,
\be\label{eqn:dhab-p-}
h_{\alpha\beta,\gamma}A'^\gamma=-\frac{8G\mu\sigma_{\alpha\beta}}{L_AA'\cdot B'/2-B'\cdot\Omega}= -\frac{8G\mu\sigma_{\alpha\beta}}{h^2+t+wx+hy}\,.
\ee
For the $+y$ diamond, the only nonvanishing contraction is
\be\label{eqn:dhab-p+}
h_{\alpha\beta,\gamma}B'^\gamma=-\frac{8G\mu\sigma_{\alpha\beta}}{L_BA'\cdot B'/2-A'\cdot\Omega}= -\frac{8G\mu\sigma_{\alpha\beta}}{h^2+t+wx-hy}\,.
\ee
For the $-x$ diamond, there is no such simplification. Equation~(\ref{eqn:dhab-A1}) becomes
\bea\label{eqn:dhab-A1-r}
h_{\alpha\beta,\gamma}&=\frac{4G\mu\sigma_{\alpha\beta}}{h^2}\left[
\frac{(1,w,-h,0)}{h^2+t+wx-hy}
+\frac{(1,w,h,0)}{h^2+t+wx+hy}\right.\\ \nonumber
&\qquad\left.- 2\frac{\left(w^2t+wx, wt+x,0,h^2z\right)} {(wt+x)^2+h^2z^2}
\right]\,.
\eea
Combining Eqs.~(\ref{eqn:dhab-p-}-\ref{eqn:dhab-A1-r}) with Eqs.~(\ref{eqn:eom-0},\ref{eqn:CS-def}), using $\sigma_{\alpha\beta}^{(-x)} A'^{(+x)\beta}= \sigma_{\alpha\beta}^{(-x)} B'^{(+x)\beta} = (2w^2, 2w, 0, 0)$ and $\sigma_{\alpha\beta}^{(-x)} A'^{(+x)\alpha}B'^{(+x)\beta} = 4w^2$, and taking $z=0$, we find the components of the acceleration for a point which sees the ring intersection,
\bleal{eqn:4acc-txy}
x^0_{,uv}&=\frac{2G\mu w^2}{h^2}\left[-\frac{2w}{x+tw}+\frac{1-h^2/w^2}{h^2+t+xw+yh}+\frac{1-h^2/w^2}{h^2+t+xw-yh}\right]\,,\\
x^1_{,uv}&=\frac{2G\mu w}{h^2}\left[\frac{2w}{x+tw}-\frac{1}{h^2+t+xw+yh}-\frac{1}{h^2+t+xw-yh}\right]\,,\\
x^2_{,uv}&=\frac{2G\mu}{h}\left[\frac{1}{h^2+t+xw+yh}-\frac{1}{h^2+t+xw-yh}\right]\,,\\
x^3_{,uv}&=0\,.
\elea

We now transport our observation point to the worldsheet of the $+x$ diamond, and so take $\Delta t=x/w$. It would be good to know where on this worldsheet feels an acceleration and where does not. The line which divides the $+x$ diamond into such regions is given by
\be\label{eqn:crit}
t=-1+\sqrt{(x-w)^2+y^2}\,,
\ee
which we call the \textit{critical line}. It is a hyperbola formed by the intersection of the $+x$ diamond's worldsheet with the future lightcone of the point on the futuremost tip of the past $+x$ diamond, which lies at $(t,x,y)=(-1,w,0)$. The critical line intersects the spacelike separated tips of the $+x$ diamond. A point in the future of the crtical line sees a ring intersection, while a point in the past of the critical line sees the figure-eight intersection.

\begin{figure}
\centerline{\includegraphics[scale=0.75]{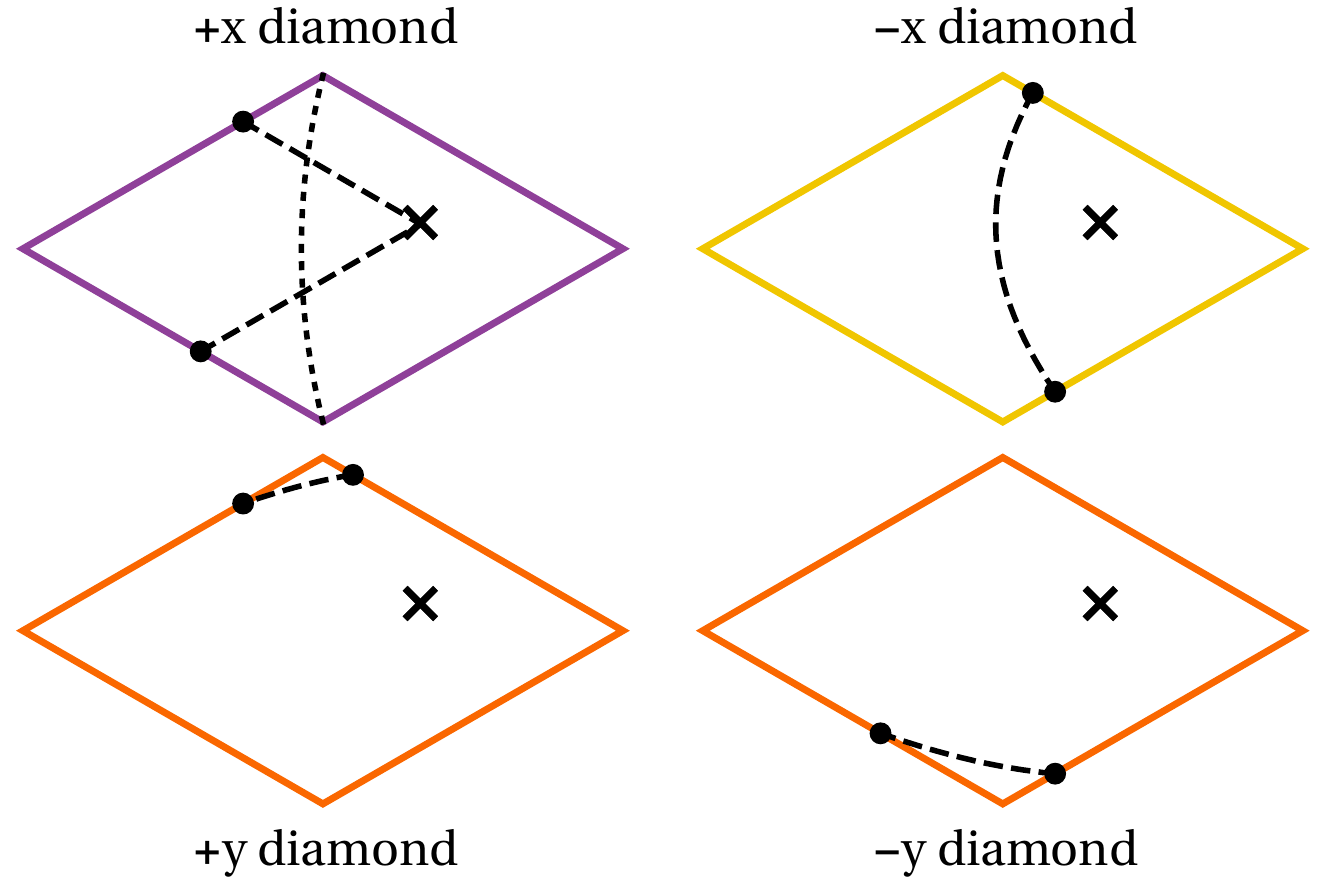}}
\caption{The intersections of the past lightcone with various diamonds for a point ${\bf\times}$ on a $+x$ diamond above the critical line (shown dotted). Solid lines indicate the worldsheet boundaries, while dashed lines indicate the intersection. Color choices are consistent with Fig.~\ref{fig:ex-ws}}\label{fig:supc-ws}
\end{figure}

\begin{figure}
\centerline{\includegraphics[scale=0.75]{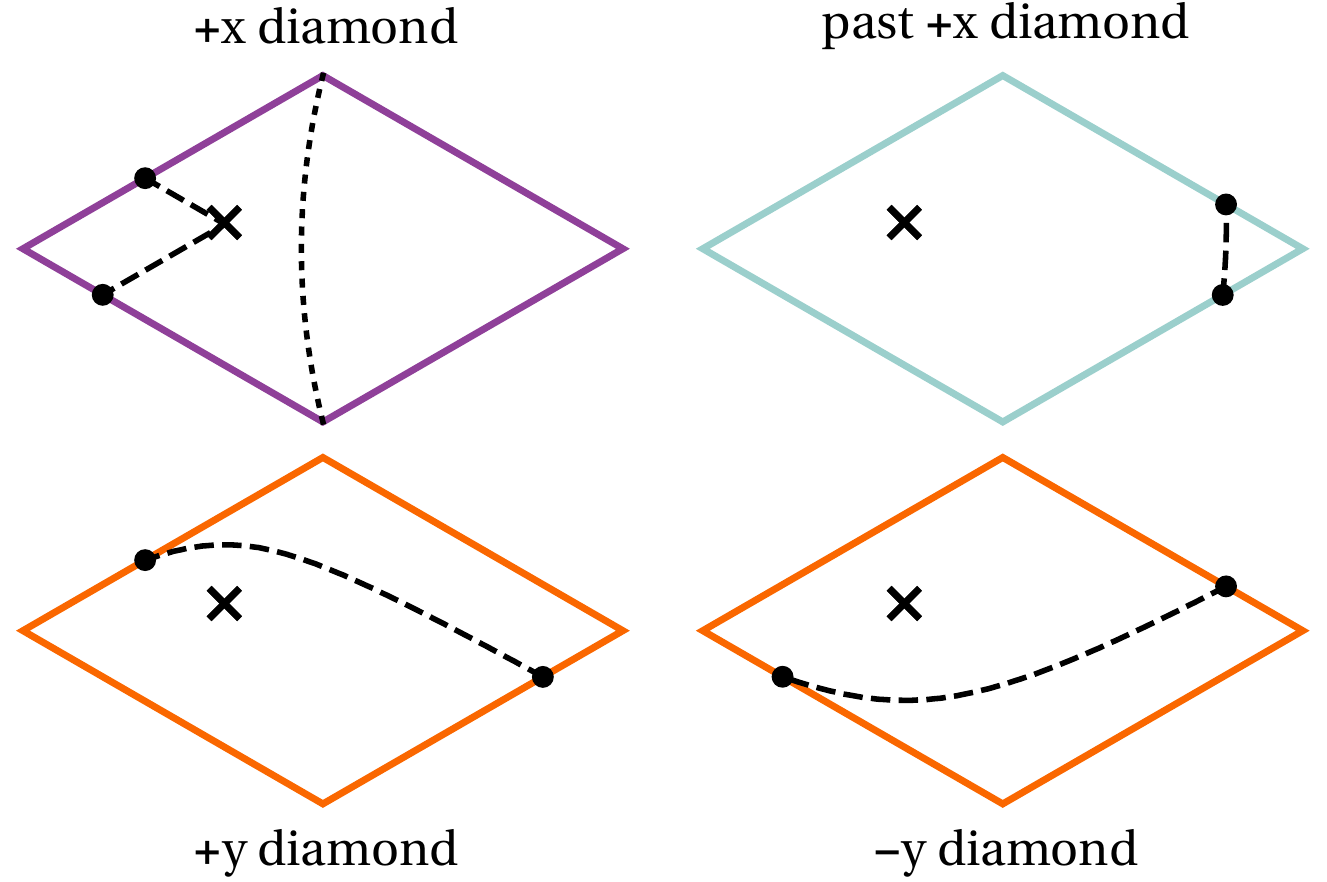}}
\caption{As in Fig.~\ref{fig:supc-ws} for an observation point ${\bf\times}$ which is below the critical line. Note that the intersection line now connects opposite edges of the $y$ diamonds, instead of connecting adjacent edges.}\label{fig:subc-ws}
\end{figure}

An example of the ring intersection for a point above the critical line is shown in Fig.~\ref{fig:supc-ws}, while those of the figure-eight intersection for a point below the critical line are shown in Fig.~\ref{fig:subc-ws}.

It is most convenient to analyze the effect of back reaction in null coordinates so we use $x=w(u+v)/2$ and $y=h(u-v)/2$, in agreement with Eq.~(\ref{eqn:x+ab}), to write \bleal{eqn:4acc-uv}
x^0_{,uv}&=\frac{2G\mu
  w^2}{h^2}\left[-\frac{2}{u+v}+\frac{1-(h/w)^2}{u+vw^2+h^2}+\frac{1-(h/w)^2}{uw^2+v+h^2}\right]\,,\label{eqn:4acc-uv0}\\ x^1_{,uv}&=\frac{2G\mu
  w}{h^2}\left[\frac{2}{u+v}-\frac{1}{u+vw^2+h^2}-\frac{1}{uw^2+v+h^2}\right]\,,\label{eqn:4acc-uv1}\\ x^2_{,uv}&=\frac{2G\mu}{h}\left[\frac{1}{u+vw^2+h^2}-\frac{1}{uw^2+v+h^2}\right]\,,\\ x^3_{,uv}&=0\,.
\elea
In Eqs.~(\ref{eqn:4acc-uv0},\ref{eqn:4acc-uv1}), the first term diverges as we approach the midline in time of the diamond, where $u = -v$. This divergence arises because source points on the $-x$ diamond come arbitrarily close to observation points on the $+x$ diamond near the time when the string intersects itself to form a double line. We can avoid the problem either by taking into account the fact that the string has some finite thickness, which will then cut off the divergence, or by perturbing the motion to allow the two parts of the string to pass some small distance from each other in the $z$ direction.

The remaining denominators in Eq.~(\ref{eqn:4acc-uv}) do not vanish for the points to which they apply.

\subsection{Modified null vectors}\label{ssec:mod-nv}

\begin{figure}
\begin{center}
\includegraphics[scale=1]{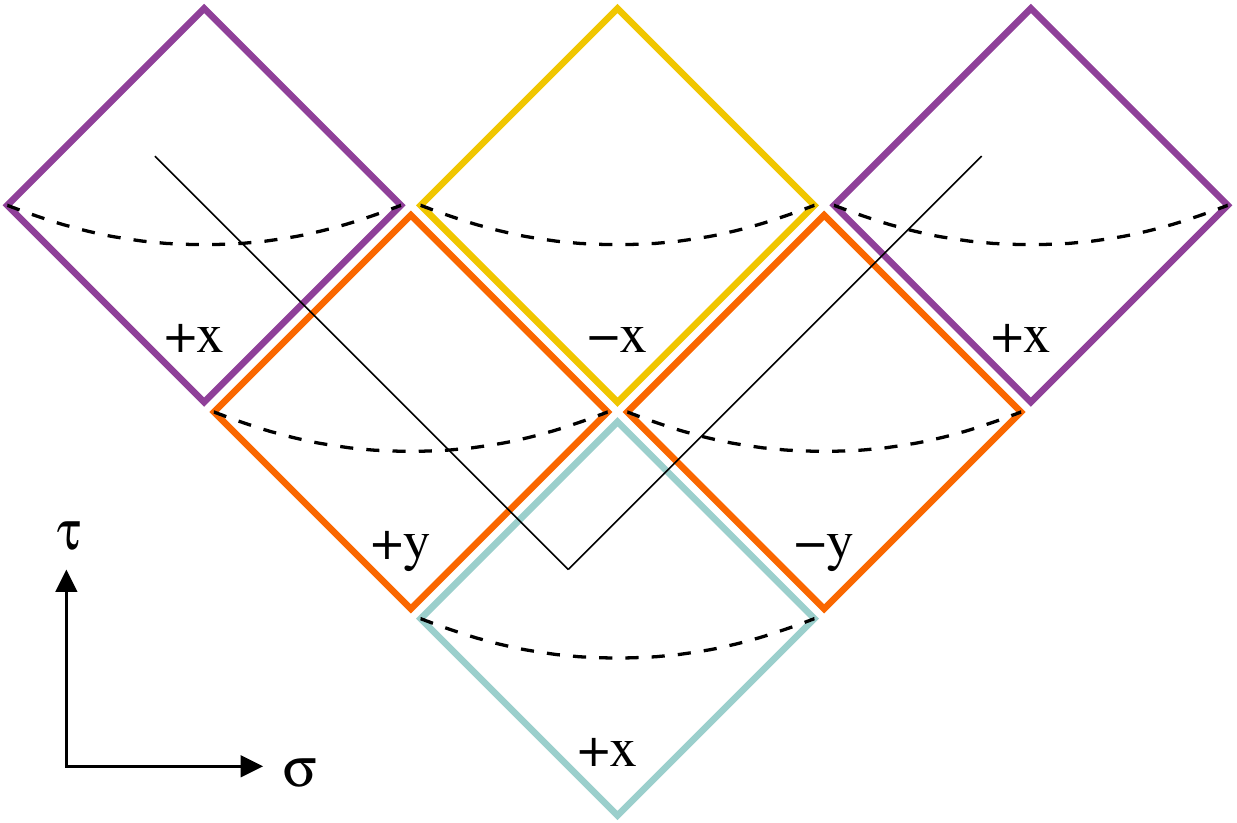}
\caption{The path through the worldsheet taken by the null (solid black) lines of a point which begins in a $+x$ diamond. The lines are drawn for one full oscillation. The color choices are consistent with Fig.~\ref{fig:ex-ws}, as the worldsheet shown in this picture is the same worldsheet, but unfolded and scaled so that all diamonds are square. Dashed black lines indicate the critical line on each diamond, below which back reaction has no effect. Adjacent diamonds in the same row point in opposite directions.  The first and third diamonds in the top row should be identified.}\label{fig:nullpath}
\end{center}
\end{figure}

To analyze the change of shape of a loop, we will consider the null tangent vectors to the worldsheet. The function $B'(\tau_+)$, for example, which has a fixed form in flat space, accumulates changes due to the presence of $x_{,uv}$. These changes are the integral of all the effects on $B'(\tau_+)$ with $\tau_+$ fixed and $\tau_-$ increasing through many oscillations \cite{Quashnock:1990wv}. For a fixed $\tau_+$ in the range $0<\tau_+<2$, the point in question moves alternately through $+x$ and $+y$ diamonds. In the case of $A'$, we fix $\tau_-$ and let $\tau_+$ vary, and for $0<\tau_-<2$ the point goes through $+x$ and $-y$ diamonds. The situation is shown in Fig.~\ref{fig:nullpath}.

In a single diamond, $u$ and $v$ range from $-1$ to $1$. But since there is no acceleration felt by points below the critical line, we only integrate $u$ over the range $(u_c,1)$ and $v$ over the range $(v_c,1)$, with the $c$ subscript indicating the value of that null parameter on the critical line. For an arbitrary initial point, the critical line value of a null parameter is found by substituting for $t$, $x$, and $y$ in Eq.~(\ref{eqn:crit}) and solving for $u$ or $v$,
\bleal{eqn:uv-critline}
u_c&=-\frac{1+vR}{R+v}\,,\\ v_c&=-\frac{1+uR}{R+u}\,,
\elea
where $R=1+2(w/h)^2$. This object has a minimum value of $R=3$ when $\phi=\pi/4$, and increases monotonically as $\phi$ decreases.

Inspection of Eq.~(\ref{eqn:4acc-uv}) tells us that there are only three kinds of integral whose solution we need to know, which we perform now.\footnote{The integral in Eq.~(\ref{eqn:pvint}) should be understood in the principal value sense. As discussed above, there is a divergence in the acceleration when the self and $-x$ diamonds intersect. We can avoid it by letting them pass by some distance $\epsilon$ in the $z$ direction. In that case the integrand will become $(u+v)/((u+v)^2+\epsilon^2)$, which will give the principal value on taking the limit as $\epsilon\to0$. No other effects of $\epsilon$ remain in this limit.}
\bleal{eqn:nv-cor-int}
\int^1_{v_c}dv\frac{1}{u+v}&=\ln\left[\frac{R+u}{1-u}\right]\,,\label{eqn:pvint}\\
\int^1_{v_c}dv\frac{1}{uw^2+v+h^2}&=\ln\left[\frac{(R'+u)(R+u)}{(1-u)^2}\right]\,,\\
\int^1_{v_c}dv\frac{1}{u+vw^2+h^2}&=\frac{1}{w^2}\ln\left[\frac{R+u}{1+u}\right]\,,
\elea
where $R'=1+2(h/w)^2$. We may recover the integrals for all other diamonds through proper exchange of $u\leftrightarrow v$ and $h\leftrightarrow w$. By applying the prefactors as given by Eq.~(\ref{eqn:4acc-uv}), and adding the effects from the $+x$ and $+y$ diamonds, we find the modification to $B'(\tau_+)$ for $0<\tau_+<2$,
\be\label{eqn:nv-cor-u}
\Delta B'^\lambda=8G\mu\left(\ln\left[\frac{(1-u)^2}{(R+u)(R'+u)}\right],\,\frac{1}{w}\ln\left[\frac{1-u}{R+u}\right],\,\frac{1}{h}\ln\left[\frac{1-u}{R'+u}\right],\,0\right)\,,
\ee
per oscillation. The modification to $A'$ is similarly
\be\label{eqn:nv-cor-v}
\Delta A'^\lambda=8G\mu\left(\ln\left[\frac{(1-v)^2}{(R+v)(R'+v)}\right],\,\frac1w\ln\left[\frac{1-v}{R+v}\right],\,-\frac1h\ln\left[\frac{1-v}{R'+v}\right],\,0\right)\,.
\ee
Equations~(\ref{eqn:nv-cor-u},\ref{eqn:nv-cor-v}) have a logarithmic divergence when then the null parameters are $+1$. This would imply that as we approach the kink, the modification to the direction of $B$ becomes larger and larger, in contradiction to what we argued in Ref.~\cite{Wachter:2016hgi}. This results from the intersection between the string and itself, which defeats the argument of Ref.~\cite{Wachter:2016hgi} that source points close to the observation point have similar motion. If we modify the loop to avoid the self intersection and instead have the strings pass by at some distance $\delta$, this distance will give a cutoff on the logarithmic divergence, so that we take $1-v$ and $1-u$ always at least $\delta$.

For the ``square'' loop where $\phi=\pi/4$, we have $h=w$ and $R' = R$, so $\Delta B'^1 = \Delta B'^2$ and similarly for $A'$. Thus there is no change to the directions of $A'$ and $B'$, and the loop retains its shape. This is a consequence of the additional symmetry in this case, as discussed in Ref.~\cite{Wachter:2016hgi}.

We have kept the parameterization of the worldsheet in terms of $\tau_\pm$. But because of the time components of Eqs.~(\ref{eqn:nv-cor-u},\ref{eqn:nv-cor-v}), the relationship between these parameters and the actual time $t$ is no longer straightforward. Nevertheless, the perturbed vectors $\tA'=A'+\Delta A'$ and $\tB'=B'+\Delta B'$ are null to first order in $G\mu$ because $A'\cdot\Delta A' = B'\cdot\Delta B'= 0$. We can integrate $\tA'$ and $\tB'$ with respect to $\tau_\pm$ to get spatially periodic functions $\tA$ and $\tB$, which have null but not unit tangent vectors. The worldsheet consists of all points which can be formed as
\be
\frac12 \left[\tA(\tau_-)+\tB(\tau_+)\right]\,.
\ee

\section{The effects of back reaction}\label{sec:eff-brxn}

\subsection{Energy loss}

When $\tau_+$ and $\tau_-$ have both increased by $T=2$, the spatial position of the string loop returns to its original form. The period of the loop is the increase in the real time $t$. To find it we integrate the time component of $\tA'$ or equivalently $\tB'$,
\be\label{eqn:tau-new}
\int_{-1}^1 dv\left\{1+8G\mu\ln\left[\frac{(1-v)^2}{(R+v)(R'+v)}\right]\right\}=2+16G\mu\left[\frac{\ln w^2}{h^2}+\frac{\ln h^2}{w^2}\right]\,.
\ee
The energy in the loop is the oscillation period times $2\mu$, so the change in energy due to back reaction is
\be\label{eqn:E-cor}
\Delta E = 32G\mu^2\left[\frac{\ln w^2}{h^2}+\frac{\ln h^2}{w^2}\right]
\ee
per oscillation. This agrees with the radiated power in Eq.~(3.9) of \cite{Garfinkle:1987yw}.

\subsection{Changes to $A$ and $B$}\label{ssec:chg-ws-fxn}

To understand the modified shape of the string, we can integrate the spatial components of Eqs.~(\ref{eqn:nv-cor-u},\ref{eqn:nv-cor-v}). We find
\bleal{eqn:ws-mod}
\Delta B^1=&\frac{8G\mu}{w}F(u,R)\\
\Delta B^2=&\frac{8G\mu}{h}F(u,R')\\
\Delta A^1=&\frac{8G\mu}{w}F(v,R)\\
\Delta A^2=&-\frac{8G\mu}{h}F(v,R')
\elea
where
\be\label{eqn:wsint-gen}
F(n,X) = \int dn\ln\left[\frac{1-n}{X+n}\right]=-(1-n)\ln(1-n)-(X+n)\ln(X+n)+\text{const.}
\ee
We will use the constant to keep the center of mass of the loop fixed. Equation~(\ref{eqn:ws-mod}) applies for $0<\tau_\pm<2$. For $2<\tau_\pm<4$, the signs of all components should be reversed and the constant chosen to keep $A$ and $B$ closed.

The total spatial change in $B$ from $\tau_+=0$ to $\tau_+=2$ is
\be
16G\mu \left[\frac{\ln w^2}{h^2}+\frac{\ln h^2}{w^2}\right]\left(w,h,0\right)\,,
\ee
which points in the direction opposite to the unmodified $B$. Thus $B$ becomes shorter by just the amount given in Eq.~(\ref{eqn:tau-new}), but does not change its overall direction. The same argument applies to $A$.

\begin{figure}
\begin{center}
\includegraphics[scale=0.75]{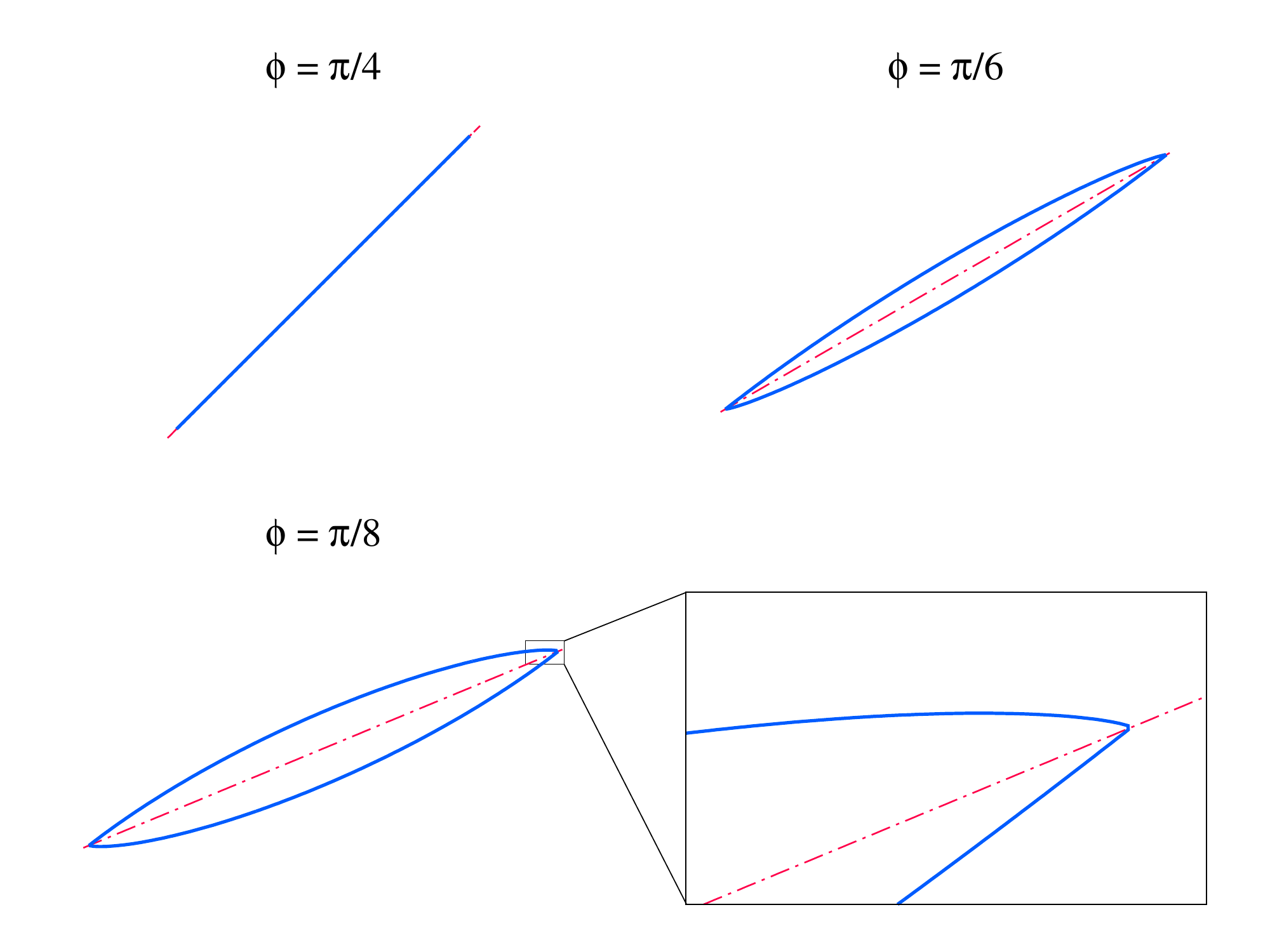}
\caption{The unmodified (dashed red) and modified (solid blue) $B$ for three different values of $\phi$. Detail of the modified $B$ as it transitions from outgoing to ingoing is given for the $\phi=\pi/8$ case. There are two general effects of back reaction: $B$ becomes shorter, and so the loop becomes shorter as well; and the turning angle of $B$ at its sharp ends becomes less than $\pi$, resulting in a ``lens'' appearance. This effect becomes weaker with increasing $\phi$, to the limit that the turning angle of the $\phi=\pi/4$ case is not affected by back reaction at all. The overall direction of $B$ (defined as the direction between the two kinks) is unchanged. We have taken $NG\mu=7\cdot10^{-3}$ in all cases.}\label{fig:Bcomp}
\end{center}
\end{figure}

We illustrate the modifications to $B$ in Fig.~\ref{fig:Bcomp}. The modifications to $A$ are given by exchange of $x\leftrightarrow-x$.

\subsection{Changes to the loop shape}\label{ssec:chg-loop}

\begin{figure}
\begin{center}
\includegraphics[scale=0.8]{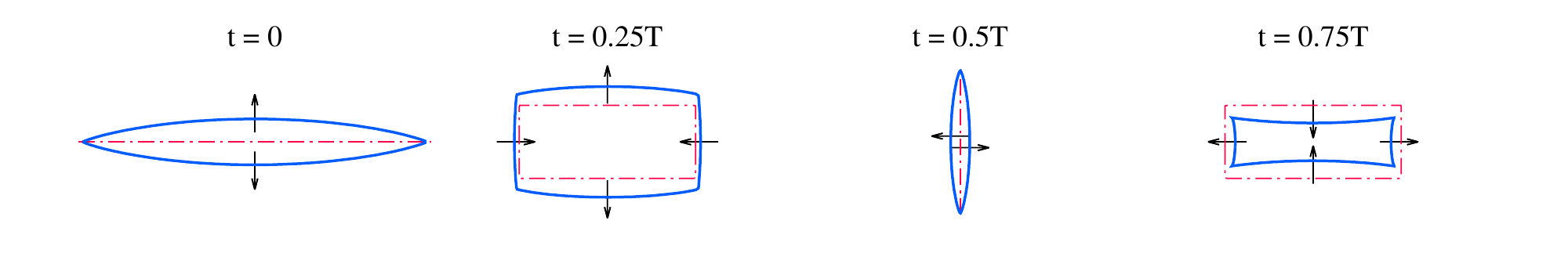}
\caption{The unmodified (dashed red) and modified (solid blue) string loops at four different constant time slices. Arrows indicate the direction of motion of a segment at a given time. The effects of back reaction are to reduce the overall size of the loop, and to introduce curvature in the kink-connecting string segments for non-square loops. It should be noted that the modified and unmodified loops have different periods, and so these slices should not be taken to be at the same coordinate times, but rather at the same fraction of the loop's oscillation. What is important is that these slightly different times are constant over the entire loop. We have taken $\phi=\pi/8$ and $NG\mu=7\cdot10^{-3}$.}\label{fig:8comp}
\end{center}
\end{figure}

Using the new forms of $A$ and $B$, we can find constant time slices of the loop's worldsheet, as shown in Fig.~\ref{fig:8comp}.
We see two effects of back reaction on non-square loops: the previously straight segments connecting kinks acquire some bend, and the loop's oscillation is no longer symmetric. In the first half of each oscillation, the kinks have an interior angle greater than $\pi/2$, and in the second half of the oscillation, less than $\pi/2$.

The modified loop at $t=0.25T$ in Fig.~\ref{fig:8comp} appears to be takes up more spatial range than the unmodified loop, even though back reaction reduces its energy. While the physical length increases, the velocities of the string segments decrease by a greater amount, such that the invariant length of the modified loop decreases.

\begin{figure}
\begin{center}
\includegraphics[scale=0.8]{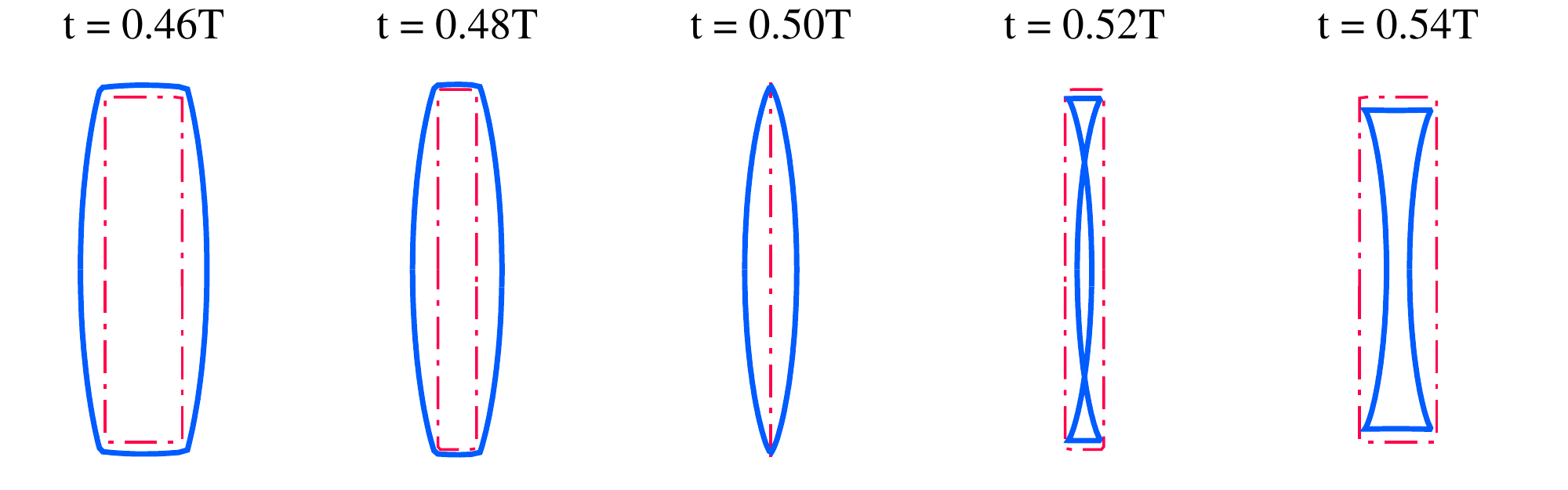}
\caption{A comparison of the unmodified (dashed red) and modified (solid blue) loop configurations just before, at, and just after the ``double line'' configuration, halfway through one oscillation. In both loops, the inwards-moving ``vertical'' segments intersect along their entire lengths. While this intersection happens all at the same time for the unmodified loop, the modified loop sees the vertical segments crossing each other progressively over some period of time. We have taken $\phi=\pi/8$ and $NG\mu=7\cdot10^{-3}$}\label{fig:8cross}
\end{center}
\end{figure}

The exact nature of the crossing of the loop segments is not as straightforward for the modified loop. The string is shown in snapshots near the double line crossing in Fig.~\ref{fig:8cross}.
The instantaneous double-line configuration has been replaced by a range of times of self-intersection. The time span between the start of subsequent self-intersections is T/2, and the range of time of self-intersection is considerably less than that.

\subsection{Cusps}

The changes to the tangent vectors $A'$ and $B'$ are also important to consider when thinking about cusp formation. For a generic loop, $A'$ and $B'$ trace out paths on the Kibble-Turok \cite{Kibble:1982cb} unit sphere. In the absence of kinks, these paths are smooth and generically intersect. These intersections produce cusps. Kinks are discontinuities in the paths of $A'$ and $B'$, which allow them to jump over each other instead of intersecting. Smoothing of kinks might lead to cusps that were not originally present.

\begin{figure}
\begin{center}
\includegraphics[scale=0.8]{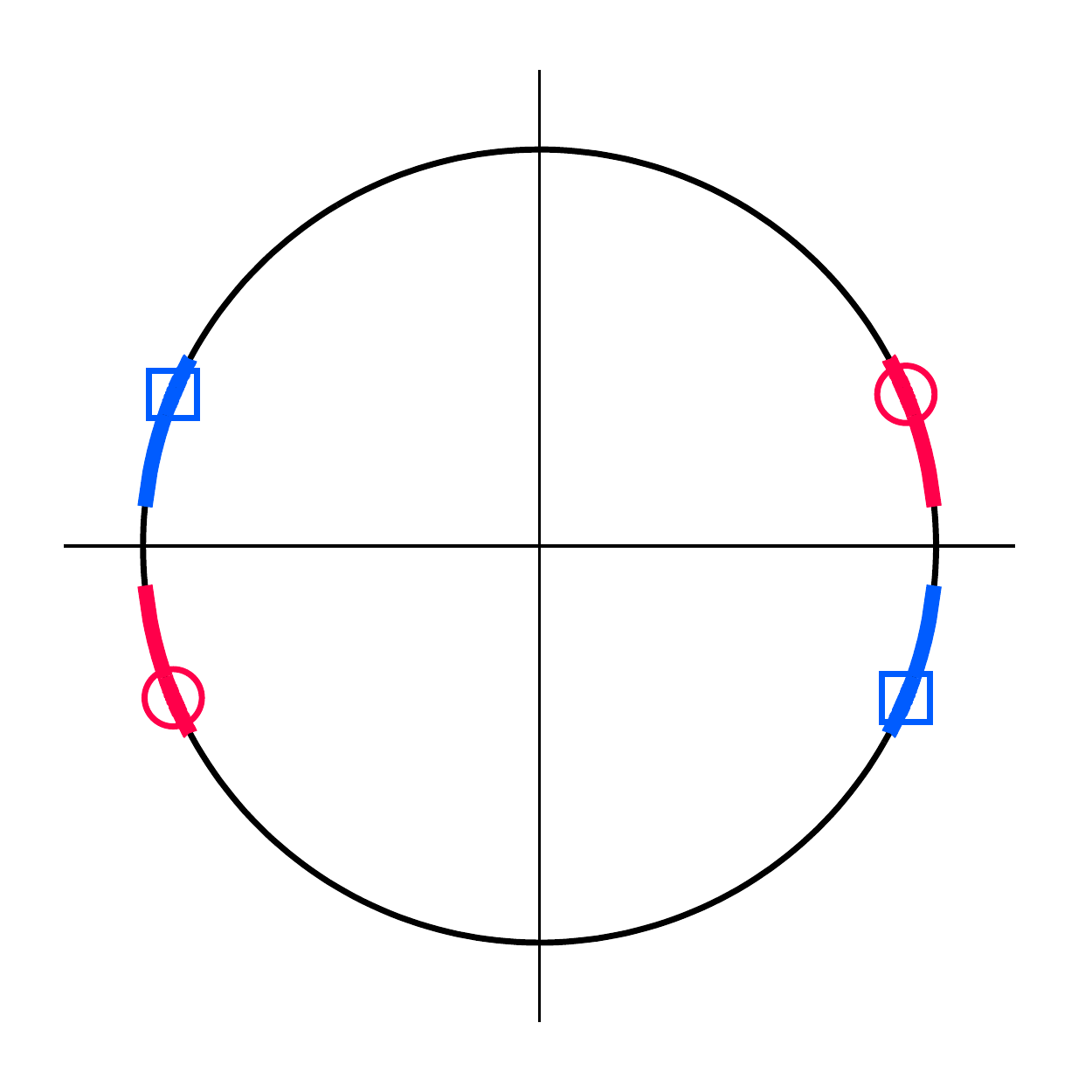}
\caption{The unmodified and modified tangent vectors on the Kibble-Turok sphere. The unmodified $A'$ and $B'$ points are marked by blue squares and red circles, respectively, while the smearing effect of back reaction is given in blue for $A'$ and red for $B'$. Back reaction has not yet acted for long enough to cause the modified tangent vectors to overlap at any point. We have chosen $\phi=\pi/8$ and $NG\mu=7\cdot10^{-3}$.}\label{fig:ucirc}
\end{center}
\end{figure}

The loops we discuss here are not generic but rather lie in a plane. Thus they always intersect themselves and so are not physically realistic. Nevertheless, they may shed some light on the question of kinks versus cusps. Because the loop lies in a plane, $A'$ and $B'$ are confined to a single great circle on the Kibble-Turok sphere. In the unmodified loop, $A'$ and $B'$ consist only of two points each, and angles $2\phi$ and $\pi-2\phi$ separate $A'$ and $B'$. Gravitational back reaction reduces the kink angles and so increases the range of directions of $A'$ and $B'$. Continued long enough, this process would cause $A'$ and $B'$ to overlap. The situation is shown in  Fig.~\ref{fig:ucirc}.

\subsection{Kink smoothing}\label{ssec:kink-sm}

We have identified two effects of back reaction, kink smoothing and loop dissipation, and we would now like to compare the relative rates at which these effects take place. 

The angle $\eta$ between the unmodified and modified direction of $B'$ can be found by taking the cross product of the spatial parts of $B'$ and $\Delta B'$,
\be
\eta \approx \sin\eta = w\Delta B'^2-h\Delta B'^1
\ee
At $u=-1$, this is 
\be\label{eqn:bend-1}
\frac{8G\mu}{hw}\ln\left[\frac{w^2}{h^2}\right]\,.
\ee
At $u=1$, there is a logarithmic divergence, as discussed above. We will modify the string so that the string passes by itself a distance $\delta$ instead of forming a double line. Then we find that $\eta$ at $u=1$ is given by
\be\label{eqn:bend1}
\frac{8G\mu}{hw}\left[w^2\ln w^2-h^2\ln h^2+(w^2-h^2)\ln(\delta/2)\right]\,.
\ee
Subtracting Eq.~(\ref{eqn:bend1}) from Eq.~(\ref{eqn:bend-1}), we find the total angle through which $B'$ turns between kinks,
\be\label{eqn:kang-cor}
\psi=\frac{8G\mu}{hw}\left[h^2\ln w^2-w^2\ln h^2-(w^2-h^2)\ln(\delta/2)\right]\,.
\ee
This is also the angle by which the original angle-$\pi$ kink in $B'$
has been straightened out.

\begin{figure}
\begin{center}
\includegraphics[scale=1]{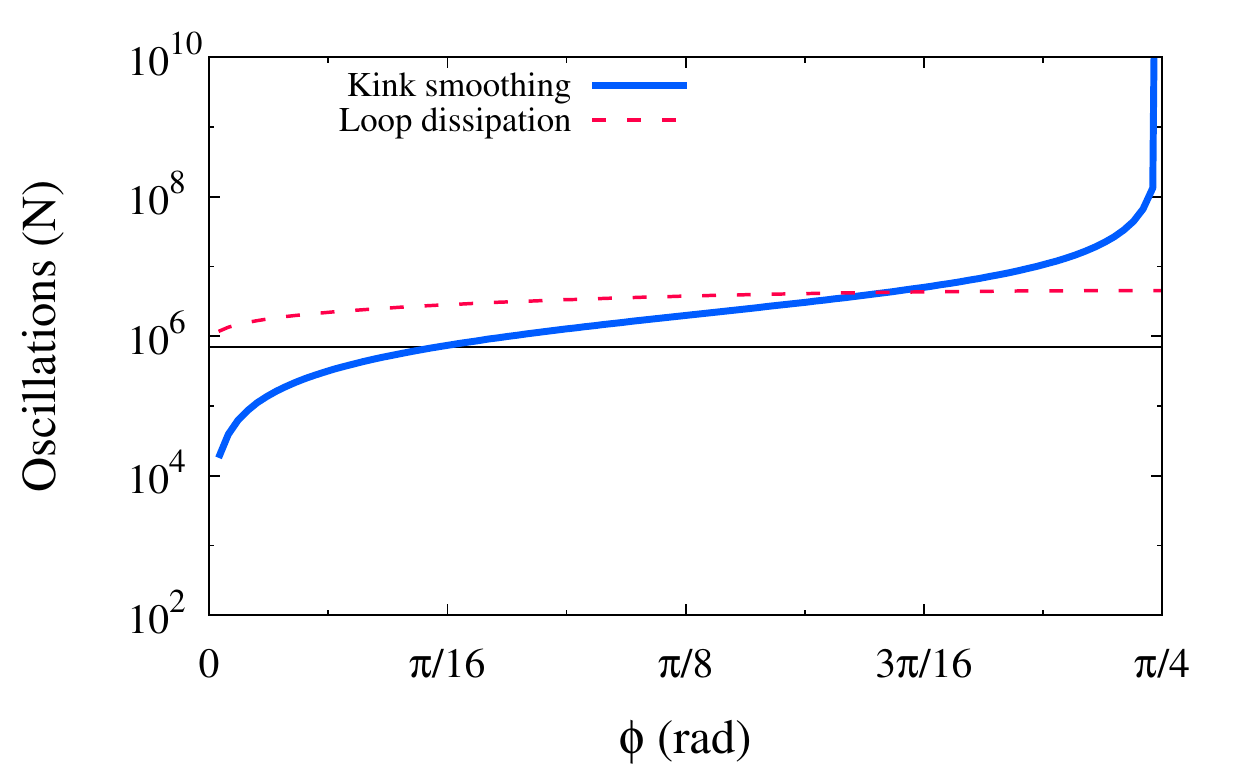}
\caption{The number of oscillations (using the first-order approximation only) for a kink to be smoothed (solid blue line) and for a loop to be dissipated (dashed red line) by back reaction. For loops close to square ($\phi=\pi/4$), kink smoothing is much slower than evaporation of the loop. For small $\phi$, kink smoothing is faster, and for a broad range in the middle the processes are roughly comparable. We have chosen $G\mu=10^{-8}$, and the black horizontal line marks our choice of $NG\mu=7\cdot10^{-3}$ for all visualizations in Sec.~\ref{sec:eff-brxn}.}\label{fig:sm-dis}
\end{center}
\end{figure}

Now, we consider a string with $G\mu=10^{-8}$ and take $\delta = 10^{-3}$ when comparing the rates of smoothing and dissipation. We are interested in comparing the timescales on which the kink angle and energy corrections become comparable to the initial angle and energy, and so we shall divide $\pi$ by $\psi$ as given by Eq.~(\ref{eqn:kang-cor}) and compare it to $4\mu$ divided by $\Delta E$ as given by Eq.~(\ref{eqn:E-cor}). These estimations provide us with an indication of the number of oscillations required to completely smooth the kink or to dissipate the loop. We compare the timescales of smoothing and dissipation in Fig.~\ref{fig:sm-dis}.
When the loop is nearly square, it mostly retains its shape over the whole evaporation process, and the number of oscillations to evaporate is meaningful. Otherwise, smoothing of the kinks takes us out of the regime in which our first-order approximations apply, so the number of oscillations should not be taken literally.

\section{Conclusions}\label{sec:conc}

A generic rectangular loop will be smoothed by back reaction in the sense that the kinks in $A$ and $B$ will become less sharp, and the segments between them will be curved, as shown in Fig.~\ref{fig:Bcomp}.  The kinks in the spatial position of the loop may become sharper or less sharp in this process, as shown in Figs.~\ref{fig:8comp} and \ref{fig:8cross}.  However, the ``square'' loop with $\phi=\pi/4$ is protected by an additional symmetry.  It does not change its geometry as it dissipates, shrinking rigidly to nothing.

The effect of smoothing accumulates more rapidly for loops which are farther from square. Loops which are nearly square will have kink angles very close to $\pi$ for the entirety of their lives, while loops which have very large aspect ratios will have their kinks significantly smoothed relatively soon after their formation.

Assuming that these effects generalize to realistic loops that do not lie in a plane, the differences in kink smoothing impact the likelihood of cusp formation.  Loops at formation have kinks but no cusps \cite{Blanco-Pillado:2015ana}.  Some of these may retain their large-angle kinks throughout their lives, and thus will never develop cusps, while on others the kinks will be rapidly opened out and cusps are likely. From Fig.~\ref{fig:sm-dis} it appears that both behaviors may be common.

Significant kink smoothing will change the loop worldsheet, and thus the backreactive acceleration, so we must account for higher-order effects to describe the entire life of a loop with precision. Combined with the difficulty of analyzing the back reaction of any but the simplest loops, this suggests that the way forward in understanding the effects of gravitational back reaction on loops lies in simulations.

\section*{Acknowledgments}

This work was supported in part by the National Science Foundation under grant number 1213930. J.M.W. would like to acknowledge the support of the John F. Burlingame Graduate Fellowship in completing this work. We thank Subir Sabharwal for helpful conversations.

\bibliography{paper}

\begin{thebibliography}{15}%
\makeatletter
\providecommand \@ifxundefined [1]{%
 \@ifx{#1\undefined}
}%
\providecommand \@ifnum [1]{%
 \ifnum #1\expandafter \@firstoftwo
 \else \expandafter \@secondoftwo
 \fi
}%
\providecommand \@ifx [1]{%
 \ifx #1\expandafter \@firstoftwo
 \else \expandafter \@secondoftwo
 \fi
}%
\providecommand \natexlab [1]{#1}%
\providecommand \enquote  [1]{``#1''}%
\providecommand \bibnamefont  [1]{#1}%
\providecommand \bibfnamefont [1]{#1}%
\providecommand \citenamefont [1]{#1}%
\providecommand \href@noop [0]{\@secondoftwo}%
\providecommand \href [0]{\begingroup \@sanitize@url \@href}%
\providecommand \@href[1]{\@@startlink{#1}\@@href}%
\providecommand \@@href[1]{\endgroup#1\@@endlink}%
\providecommand \@sanitize@url [0]{\catcode `\\12\catcode `\$12\catcode
  `\&12\catcode `\#12\catcode `\^12\catcode `\_12\catcode `\%12\relax}%
\providecommand \@@startlink[1]{}%
\providecommand \@@endlink[0]{}%
\providecommand \url  [0]{\begingroup\@sanitize@url \@url }%
\providecommand \@url [1]{\endgroup\@href {#1}{\urlprefix }}%
\providecommand \urlprefix  [0]{URL }%
\providecommand \Eprint [0]{\href }%
\providecommand \doibase [0]{http://dx.doi.org/}%
\providecommand \selectlanguage [0]{\@gobble}%
\providecommand \bibinfo  [0]{\@secondoftwo}%
\providecommand \bibfield  [0]{\@secondoftwo}%
\providecommand \translation [1]{[#1]}%
\providecommand \BibitemOpen [0]{}%
\providecommand \bibitemStop [0]{}%
\providecommand \bibitemNoStop [0]{.\EOS\space}%
\providecommand \EOS [0]{\spacefactor3000\relax}%
\providecommand \BibitemShut  [1]{\csname bibitem#1\endcsname}%
\let\auto@bib@innerbib\@empty
\bibitem [{\citenamefont {Jeannerot}\ \emph {et~al.}(2003)\citenamefont
  {Jeannerot}, \citenamefont {Rocher},\ and\ \citenamefont
  {Sakellariadou}}]{Jeannerot:2003qv}%
  \BibitemOpen
  \bibfield  {author} {\bibinfo {author} {\bibfnamefont {Rachel}\ \bibnamefont
  {Jeannerot}}, \bibinfo {author} {\bibfnamefont {Jonathan}\ \bibnamefont
  {Rocher}}, \ and\ \bibinfo {author} {\bibfnamefont {Mairi}\ \bibnamefont
  {Sakellariadou}},\ }\bibfield  {title} {\enquote {\bibinfo {title} {{How
  generic is cosmic string formation in SUSY GUTs}},}\ }\href {\doibase
  10.1103/PhysRevD.68.103514} {\bibfield  {journal} {\bibinfo  {journal} {Phys.
  Rev.}\ }\textbf {\bibinfo {volume} {D68}},\ \bibinfo {pages} {103514}
  (\bibinfo {year} {2003})},\ \Eprint {http://arxiv.org/abs/hep-ph/0308134}
  {arXiv:hep-ph/0308134 [hep-ph]} \BibitemShut {NoStop}%
\bibitem [{\citenamefont {Sarangi}\ and\ \citenamefont
  {Tye}(2002)}]{Sarangi:2002yt}%
  \BibitemOpen
  \bibfield  {author} {\bibinfo {author} {\bibfnamefont {Saswat}\ \bibnamefont
  {Sarangi}}\ and\ \bibinfo {author} {\bibfnamefont {S.~H.~Henry}\ \bibnamefont
  {Tye}},\ }\bibfield  {title} {\enquote {\bibinfo {title} {{Cosmic string
  production towards the end of brane inflation}},}\ }\href {\doibase
  10.1016/S0370-2693(02)01824-5} {\bibfield  {journal} {\bibinfo  {journal}
  {Phys. Lett.}\ }\textbf {\bibinfo {volume} {B536}},\ \bibinfo {pages}
  {185--192} (\bibinfo {year} {2002})},\ \Eprint
  {http://arxiv.org/abs/hep-th/0204074} {arXiv:hep-th/0204074 [hep-th]}
  \BibitemShut {NoStop}%
\bibitem [{\citenamefont {Dvali}\ and\ \citenamefont
  {Vilenkin}(2004)}]{Dvali:2003zj}%
  \BibitemOpen
  \bibfield  {author} {\bibinfo {author} {\bibfnamefont {Gia}\ \bibnamefont
  {Dvali}}\ and\ \bibinfo {author} {\bibfnamefont {Alexander}\ \bibnamefont
  {Vilenkin}},\ }\bibfield  {title} {\enquote {\bibinfo {title} {{Formation and
  evolution of cosmic D strings}},}\ }\href {\doibase
  10.1088/1475-7516/2004/03/010} {\bibfield  {journal} {\bibinfo  {journal}
  {JCAP}\ }\textbf {\bibinfo {volume} {0403}},\ \bibinfo {pages} {010}
  (\bibinfo {year} {2004})},\ \Eprint {http://arxiv.org/abs/hep-th/0312007}
  {arXiv:hep-th/0312007 [hep-th]} \BibitemShut {NoStop}%
\bibitem [{\citenamefont {Vilenkin}\ and\ \citenamefont
  {Shellard}(2000)}]{Vilenkin:2000jqa}%
  \BibitemOpen
  \bibfield  {author} {\bibinfo {author} {\bibfnamefont {A.}~\bibnamefont
  {Vilenkin}}\ and\ \bibinfo {author} {\bibfnamefont {E.~P.~S.}\ \bibnamefont
  {Shellard}},\ }\href
  {http://www.cambridge.org/mw/academic/subjects/physics/theoretical-physics-and-mathematical-physics/cosmic-strings-and-other-topological-defects?format=PB}
  {\emph {\bibinfo {title} {{Cosmic Strings and Other Topological Defects}}}}\
  (\bibinfo  {publisher} {Cambridge University Press},\ \bibinfo {year}
  {2000})\BibitemShut {NoStop}%
\bibitem [{\citenamefont {Damour}\ and\ \citenamefont
  {Vilenkin}(2001)}]{Damour:2001bk}%
  \BibitemOpen
  \bibfield  {author} {\bibinfo {author} {\bibfnamefont {Thibault}\
  \bibnamefont {Damour}}\ and\ \bibinfo {author} {\bibfnamefont {Alexander}\
  \bibnamefont {Vilenkin}},\ }\bibfield  {title} {\enquote {\bibinfo {title}
  {{Gravitational wave bursts from cusps and kinks on cosmic strings}},}\
  }\href {\doibase 10.1103/PhysRevD.64.064008} {\bibfield  {journal} {\bibinfo
  {journal} {Phys. Rev.}\ }\textbf {\bibinfo {volume} {D64}},\ \bibinfo {pages}
  {064008} (\bibinfo {year} {2001})},\ \Eprint
  {http://arxiv.org/abs/gr-qc/0104026} {arXiv:gr-qc/0104026 [gr-qc]}
  \BibitemShut {NoStop}%
\bibitem [{\citenamefont {Cai}\ \emph {et~al.}(2012)\citenamefont {Cai},
  \citenamefont {Sabancilar}, \citenamefont {Steer},\ and\ \citenamefont
  {Vachaspati}}]{Cai:2012zd}%
  \BibitemOpen
  \bibfield  {author} {\bibinfo {author} {\bibfnamefont {Yi-Fu}\ \bibnamefont
  {Cai}}, \bibinfo {author} {\bibfnamefont {Eray}\ \bibnamefont {Sabancilar}},
  \bibinfo {author} {\bibfnamefont {Daniele~A.}\ \bibnamefont {Steer}}, \ and\
  \bibinfo {author} {\bibfnamefont {Tanmay}\ \bibnamefont {Vachaspati}},\
  }\bibfield  {title} {\enquote {\bibinfo {title} {{Radio Broadcasts from
  Superconducting Strings}},}\ }\href {\doibase 10.1103/PhysRevD.86.043521}
  {\bibfield  {journal} {\bibinfo  {journal} {Phys. Rev.}\ }\textbf {\bibinfo
  {volume} {D86}},\ \bibinfo {pages} {043521} (\bibinfo {year} {2012})},\
  \Eprint {http://arxiv.org/abs/1205.3170} {arXiv:1205.3170 [astro-ph.CO]}
  \BibitemShut {NoStop}%
\bibitem [{\citenamefont {Berezinsky}\ \emph {et~al.}(2011)\citenamefont
  {Berezinsky}, \citenamefont {Sabancilar},\ and\ \citenamefont
  {Vilenkin}}]{Berezinsky:2011cp}%
  \BibitemOpen
  \bibfield  {author} {\bibinfo {author} {\bibfnamefont {Veniamin}\
  \bibnamefont {Berezinsky}}, \bibinfo {author} {\bibfnamefont {Eray}\
  \bibnamefont {Sabancilar}}, \ and\ \bibinfo {author} {\bibfnamefont
  {Alexander}\ \bibnamefont {Vilenkin}},\ }\bibfield  {title} {\enquote
  {\bibinfo {title} {{Extremely High Energy Neutrinos from Cosmic Strings}},}\
  }\href {\doibase 10.1103/PhysRevD.84.085006} {\bibfield  {journal} {\bibinfo
  {journal} {Phys. Rev.}\ }\textbf {\bibinfo {volume} {D84}},\ \bibinfo {pages}
  {085006} (\bibinfo {year} {2011})},\ \Eprint {http://arxiv.org/abs/1108.2509}
  {arXiv:1108.2509 [astro-ph.CO]} \BibitemShut {NoStop}%
\bibitem [{\citenamefont {Blanco-Pillado}\ \emph {et~al.}(2014)\citenamefont
  {Blanco-Pillado}, \citenamefont {Olum},\ and\ \citenamefont
  {Shlaer}}]{Blanco-Pillado:2013qja}%
  \BibitemOpen
  \bibfield  {author} {\bibinfo {author} {\bibfnamefont {Jose~J.}\ \bibnamefont
  {Blanco-Pillado}}, \bibinfo {author} {\bibfnamefont {Ken~D.}\ \bibnamefont
  {Olum}}, \ and\ \bibinfo {author} {\bibfnamefont {Benjamin}\ \bibnamefont
  {Shlaer}},\ }\bibfield  {title} {\enquote {\bibinfo {title} {{The number of
  cosmic string loops}},}\ }\href {\doibase 10.1103/PhysRevD.89.023512}
  {\bibfield  {journal} {\bibinfo  {journal} {Phys. Rev.}\ }\textbf {\bibinfo
  {volume} {D89}},\ \bibinfo {pages} {023512} (\bibinfo {year} {2014})},\
  \Eprint {http://arxiv.org/abs/1309.6637} {arXiv:1309.6637 [astro-ph.CO]}
  \BibitemShut {NoStop}%
\bibitem [{\citenamefont {Aasi}\ \emph {et~al.}(2014)\citenamefont {Aasi} \emph
  {et~al.}}]{Aasi:2013vna}%
  \BibitemOpen
  \bibfield  {author} {\bibinfo {author} {\bibfnamefont {J.}~\bibnamefont
  {Aasi}} \emph {et~al.} (\bibinfo {collaboration} {VIRGO, LIGO Scientific}),\
  }\bibfield  {title} {\enquote {\bibinfo {title} {{Constraints on cosmic
  strings from the LIGO-Virgo gravitational-wave detectors}},}\ }\href
  {\doibase 10.1103/PhysRevLett.112.131101} {\bibfield  {journal} {\bibinfo
  {journal} {Phys. Rev. Lett.}\ }\textbf {\bibinfo {volume} {112}},\ \bibinfo
  {pages} {131101} (\bibinfo {year} {2014})},\ \Eprint
  {http://arxiv.org/abs/1310.2384} {arXiv:1310.2384 [gr-qc]} \BibitemShut
  {NoStop}%
\bibitem [{\citenamefont {Blanco-Pillado}\ \emph {et~al.}(2015)\citenamefont
  {Blanco-Pillado}, \citenamefont {Olum},\ and\ \citenamefont
  {Shlaer}}]{Blanco-Pillado:2015ana}%
  \BibitemOpen
  \bibfield  {author} {\bibinfo {author} {\bibfnamefont {Jose~J.}\ \bibnamefont
  {Blanco-Pillado}}, \bibinfo {author} {\bibfnamefont {Ken~D.}\ \bibnamefont
  {Olum}}, \ and\ \bibinfo {author} {\bibfnamefont {Benjamin}\ \bibnamefont
  {Shlaer}},\ }\bibfield  {title} {\enquote {\bibinfo {title} {{Cosmic string
  loop shapes}},}\ }\href {\doibase 10.1103/PhysRevD.92.063528} {\bibfield
  {journal} {\bibinfo  {journal} {Phys. Rev.}\ }\textbf {\bibinfo {volume}
  {D92}},\ \bibinfo {pages} {063528} (\bibinfo {year} {2015})},\ \Eprint
  {http://arxiv.org/abs/1508.02693} {arXiv:1508.02693 [astro-ph.CO]}
  \BibitemShut {NoStop}%
\bibitem [{\citenamefont {Allen}\ and\ \citenamefont
  {Ottewill}(2001)}]{Allen:2000ia}%
  \BibitemOpen
  \bibfield  {author} {\bibinfo {author} {\bibfnamefont {Bruce}\ \bibnamefont
  {Allen}}\ and\ \bibinfo {author} {\bibfnamefont {Adrian~C.}\ \bibnamefont
  {Ottewill}},\ }\bibfield  {title} {\enquote {\bibinfo {title} {{Wave forms
  for gravitational radiation from cosmic string loops}},}\ }\href {\doibase
  10.1103/PhysRevD.63.063507} {\bibfield  {journal} {\bibinfo  {journal} {Phys.
  Rev.}\ }\textbf {\bibinfo {volume} {D63}},\ \bibinfo {pages} {063507}
  (\bibinfo {year} {2001})},\ \Eprint {http://arxiv.org/abs/gr-qc/0009091}
  {arXiv:gr-qc/0009091 [gr-qc]} \BibitemShut {NoStop}%
\bibitem [{\citenamefont {Garfinkle}\ and\ \citenamefont
  {Vachaspati}(1987)}]{Garfinkle:1987yw}%
  \BibitemOpen
  \bibfield  {author} {\bibinfo {author} {\bibfnamefont {David}\ \bibnamefont
  {Garfinkle}}\ and\ \bibinfo {author} {\bibfnamefont {Tanmay}\ \bibnamefont
  {Vachaspati}},\ }\bibfield  {title} {\enquote {\bibinfo {title} {Radiation
  from kinky, cuspless cosmic loops},}\ }\href {\doibase
  10.1103/PhysRevD.36.2229} {\bibfield  {journal} {\bibinfo  {journal} {Phys.
  Rev. D}\ }\textbf {\bibinfo {volume} {36}},\ \bibinfo {pages} {2229--2241}
  (\bibinfo {year} {1987})}\BibitemShut {NoStop}%
\bibitem [{\citenamefont {Quashnock}\ and\ \citenamefont
  {Spergel}(1990)}]{Quashnock:1990wv}%
  \BibitemOpen
  \bibfield  {author} {\bibinfo {author} {\bibfnamefont {Jean~M.}\ \bibnamefont
  {Quashnock}}\ and\ \bibinfo {author} {\bibfnamefont {David~N.}\ \bibnamefont
  {Spergel}},\ }\bibfield  {title} {\enquote {\bibinfo {title} {{Gravitational
  Selfinteractions of Cosmic Strings}},}\ }\href {\doibase
  10.1103/PhysRevD.42.2505} {\bibfield  {journal} {\bibinfo  {journal}
  {Phys.Rev.}\ }\textbf {\bibinfo {volume} {D42}},\ \bibinfo {pages}
  {2505--2520} (\bibinfo {year} {1990})}\BibitemShut {NoStop}%
\bibitem [{\citenamefont {Wachter}\ and\ \citenamefont
  {Olum}(2016)}]{Wachter:2016hgi}%
  \BibitemOpen
  \bibfield  {author} {\bibinfo {author} {\bibfnamefont {Jeremy~M.}\
  \bibnamefont {Wachter}}\ and\ \bibinfo {author} {\bibfnamefont {Ken~D.}\
  \bibnamefont {Olum}},\ }\bibfield  {title} {\enquote {\bibinfo {title}
  {{Gravitational smoothing of kinks on cosmic string loops}},}\ }\href@noop {}
  {\  (\bibinfo {year} {2016})},\ \Eprint {http://arxiv.org/abs/1609.01153}
  {arXiv:1609.01153 [gr-qc]} \BibitemShut {NoStop}%
\bibitem [{\citenamefont {Kibble}\ and\ \citenamefont
  {Turok}(1982)}]{Kibble:1982cb}%
  \BibitemOpen
  \bibfield  {author} {\bibinfo {author} {\bibfnamefont {T.~W.~B.}\
  \bibnamefont {Kibble}}\ and\ \bibinfo {author} {\bibfnamefont {Neil}\
  \bibnamefont {Turok}},\ }\bibfield  {title} {\enquote {\bibinfo {title}
  {{Selfintersection of Cosmic Strings}},}\ }\href {\doibase
  10.1016/0370-2693(82)90993-5} {\bibfield  {journal} {\bibinfo  {journal}
  {Phys. Lett.}\ }\textbf {\bibinfo {volume} {B116}},\ \bibinfo {pages}
  {141--143} (\bibinfo {year} {1982})}\BibitemShut {NoStop}%
\end{thebibliography}%

\end{document}